\documentclass[12pt]{article}
\usepackage{epsfig}

\date{}

\def\Journal#1#2#3#4{{#1} {\bf #2} (#4) #3}

\def\NCA{\em Nuovo Cimento}

\def\NPB{{\em Nucl. Phys.} B}
\def\PLB{{\em Phys. Lett.} B} 
\def\PRL{\em Phys. Rev. Lett.} 
\def\PRD{{\em Phys. Rev.} D} 
\def\ZPC{{\em Z. Phys.} C} 
\def\PR{\em Phys. Rev.}
\def\MPLA{{\em Mod. Phys. Lett.} A} 
\def\SPU{\em Sov. Phys. Usp.}
\def\PRC{{\em Phys. Rep.} C}
\def\IJMPA{{\em Int. J. Mod. Phys.} A}

\newcommand{\lwig}{\mbox{\,\raisebox{.3ex}
    {$<$}$\!\!\!\!\!$\raisebox{-.9ex}{$\sim$}\,}}
\newcommand{\gwig}{\mbox{\,\raisebox{.3ex}
    {$>$}$\!\!\!\!\!$\raisebox{-.9ex}{$\sim$}}\,}

\newcommand{\ai}{{\overline{I}}} 
\newcommand{\iai}{I\overline{I}}

\newcommand{\ii}{{\rm i}} 
 
\newcommand{\xpr}{{x^\prime}}
 
\newcommand{\xbj}{x_{\rm Bj}}
\newcommand{\ybj}{y_{\rm Bj}} 

\newcommand{\Qprime}{{Q^\prime}}

\begin{document}
\title{{\normalsize\rightline{DESY 96-203}\rightline{hep-ph/9610213}} 
\vskip 1cm 
      \bf Instanton-Induced Processes\\
          in Deep-Inelastic Scattering\thanks{To be published in:
          {\it Quarks `96}, Proc. IXth International Seminar, 
          Yaroslavl, Russia, May 5-11, 1996.}
                          \\
  \vspace{11mm}}
\author{A. Ringwald and F. Schrempp\\[3mm] Deutsches
  Elektronen-Synchrotron DESY, Hamburg, Germany}
\begin{titlepage} 
  \maketitle
\begin{abstract}
We present a status report of our systematic theoretical and phenomenological 
study of QCD-instanton induced processes in deep-inelastic scattering. We show
that this regime plays a distinguished r\^{o}le for studying manifestations of 
QCD-instantons, since the typical hard momentum scale ${\mathcal Q}$ provides 
a dynamical infrared cutoff for the instanton size 
$\rho\lwig {\mathcal O}(1/{\mathcal Q})$. For deep-inelastic 
scattering at HERA, we present a preliminary theoretical 
estimate of the total instanton-induced cross-section (subject to appropriate
kinematical cuts). It is surprisingly large, in the ${\mathcal O}(1-100)$ pb 
range, albeit still uncertain. We report on our investigation of the discovery 
potential for instanton-induced events at HERA by means of a Monte Carlo event
generator. It is based on a detailed study of the characteristic signatures of
the final state, like a large total transverse energy, 
$E_{T}={\mathcal O}(20)$ GeV, a large multiplicity, $n={\mathcal O}(25)$, and 
a flavour-democratic production of hadrons. A combination of event shape 
information with searches of $K^{0}$ mesons, muons, and multiplicity cuts 
might help to discriminate further the QCD-instanton induced processes from 
the standard perturbative QCD background.
\end{abstract} 
\thispagestyle{empty}
\end{titlepage}
\newpage \setcounter{page}{2}

\section{Introduction}

The Standard Model of strong (QCD) and  electro-weak (QFD) interactions  
is remarkably successful. Its perturbative formulation appears to be 
theoretically consistent and agrees with present experiments. 
Yet, the existence of Adler-Bell-Jackiw anomalies~\cite{abj} 
implies that there are also processes that cannot be described by  
conventional perturbation theory~\cite{th}.  They give rise to a violation
of certain fermionic quantum numbers, notably  chirality ($Q_{5}$) in 
(massless) QCD and $B+L$ in QFD.

Such anomalous processes are induced by {\it instantons}~\cite{bpst} 
which represent tunnelling processes in Yang-Mills
gauge theories, associated with the highly degenerate 
vacuum structure~\cite{tunnel}.

An experimental discovery of such a novel, 
non-perturbative manifestation of non-abelian gauge theories would 
clearly be of basic significance.

A number of results has revived the interest in instanton-induced processes
during recent years: 
\begin{itemize}
\item First of all, it was shown~\cite{r,m} that the generic exponential 
suppression of these tunnelling rates, $\propto \exp (-4\pi /\alpha )$, may be
overcome at {\it high energies}, mainly due to multi-gauge boson emission
in addition to the minimally required fermionic final state.
 
\item A pioneering and encouraging theoretical estimate 
of the size of the instanton ($I$) induced contribution 
to the gluon structure functions in deep-inelastic scattering (DIS) 
was recently performed in Ref.~\cite{bb}. It was argued 
that it is possible to isolate a well-defined and sizable
instanton contribution in the regime of small QCD-gauge coupling, on 
account of the (large) photon virtuality $Q^2$. While the instanton-induced
contribution to the gluon structure functions turned 
out to be small at larger values of the Bjorken variable $x$,
it was found in Ref.~\cite{bb} to increase dramatically towards 
smaller $x$.

\item Last not least, a systematic phenomenological and theoretical study
is under way~\cite{rs,grs,rs1,ggmrs,mrs1}, which  clearly indicates that 
deep-inelastic $e p$ scattering at HERA now offers a unique window to 
experimentally detect QCD-instanton induced processes through their 
characteristic final-state signature. The searches for instanton-induced 
events have just started at HERA and a first upper limit of $0.9$ nb 
at $95\%$ confidence level for the cross-section of QCD-instanton induced 
events has been placed by the H1 Collaboration~\cite{H1}. New, improved 
search strategies are being developped~\cite{ggmrs} with the help of a 
Monte Carlo generator (QCDINS 1.3)~\cite{grs} for instanton-induced events. 
\end{itemize}

In this review, we report on the present status of our 
broad and systematic study of QCD-instanton induced processes
in deep-inelastic scattering~\cite{rs,grs,rs1,ggmrs,mrs1,mrs2,grs2}.

In Sect.~\ref{whatisspecial}, we start by summarizing the important 
theoretical result~\cite{mrs1} that instanton-induced processes in 
deep-inelastic scattering do not suffer from the usually encountered 
infrared (IR) divergencies
associated with the integration over the instanton size. In fact, 
the inverse hard momentum scale in DIS provides a dynamical IR cutoff
for the instanton size. We point out the close analogy of QCD-instanton
induced processes in DIS to those in QFD.  
In Sect.~\ref{sec:struc}, we present the current status of our
ongoing theoretical study~\cite{mrs2} of the 
$I$-induced contribution to deep-inelastic lepto-production. 
Of interest is, of course, the first (preliminary)
estimate of the total $I$-induced cross-section at HERA.   
 Section~\ref{sec:monte} is devoted to our phenomenological investigation of  
the discovery potential for instanton-induced events at HERA 
by means of a Monte Carlo event generator~\cite{grs,grs2}.
The emphasis rests on the
study of  the characteristic signatures of the final state.
We also report on the first searches of instanton-induced events in DIS at
HERA.
Finally, we discuss improved search strategies, which can serve to enhance
the signal from $I$-induced events relative to the background from normal 
DIS events.   
 
\begin{figure} 
\begin{center}
  \epsfig{file=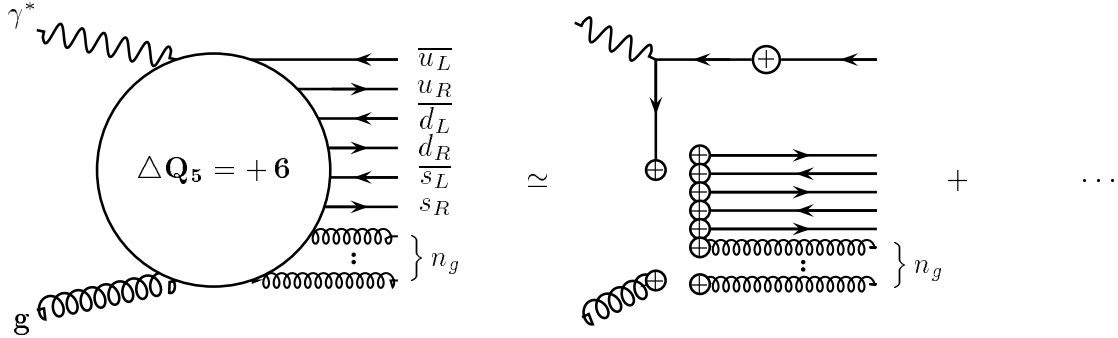,width=15cm}
\caption[dum]{\label{famplgen}
  Instanton-induced {\it chirality-violating} process,\hfill\\ $\gamma^{\ast}
  +{\rm g}\rightarrow \sum_{\rm flavours}^{n_{f}}\left[ \overline{{\rm 
      q}_{L}} +{\rm q}_{R}\right] +n_{g}\,{\rm g}$, corresponding to 
   three massless flavours 
  ($n_{f}=3$). }
 \end{center}
\end{figure}

\section{\label{whatisspecial} What is Special in Deep-Inelastic Scattering?}

As is well known, instanton calculations in QCD are generically
plagued by IR divergencies associated with the integration over the
instanton size~\cite{irdiv}. In this Section, we shall demonstrate explicitly 
that the typical hard momentum scale ${\mathcal Q}$ 
in deep-inelastic scattering provides a dynamical IR cutoff for the
instanton size $\rho_I\lwig{\mathcal O}(1/{\mathcal Q})$, such that the
integration over the size parameter is {\it finite}~\cite{mrs1}. 
Let us consider the amplitude for the relevant instanton-induced 
{\it chirality-violating} photon-gluon process (see Fig.~\ref{famplgen}), 
\begin{equation}
\label{instproc}
\gamma^\ast + {\rm g}\ \Rightarrow 
\sum_{\rm flavours}^{n_{f}}\left[\overline{{\rm q}_L}+{\rm q}_R\right] 
        + n_{\rm g}\,{\rm g}
\ ;\hspace{4ex}
\left(\triangle Q_5 \equiv \triangle (Q_R-Q_L)= 2\,n_f\right) \, .
\end{equation}
The strategy is  to first set up  the respective Green's function 
according to  standard instanton-perturbation theory in Euclidean 
space~\cite{th,bccl,ber,abc,r}, then to Fourier transform 
to momentum space, to amputate the external legs, and finally to continue
the result to Minkowski space. 
The basic building blocks are (in Euclidean space
and in the singular gauge):  

\bigskip
\noindent
i) The classical instanton gauge field\footnote{\label{footsigma}
We use the standard notations, in Euclidean space: 
$\sigma_\mu =(-\ii\,\vec{\sigma},1)$, $\overline{\sigma}_\mu =
(\ii\,\vec{\sigma},1)$, and in Minkowski space:
$\sigma_\mu =(1,\vec{\sigma})$, $\overline{\sigma}_\mu =(1,-\vec{\sigma})$, 
where $\vec{\sigma}$ are the Pauli
matrices. Furthermore, for any
four-vector $v_\mu $, we use the shorthand  
$v \equiv v\cdot \sigma,\ \overline{v} \equiv v\cdot \overline{\sigma}$.} 
$A^{(I)}_{\mu ^{\prime}}$~\cite{bpst},
\begin{eqnarray}
A_{\mu ^{\prime}}^{(I)}(x)&=&-\ii\,\frac{2\,\pi^{2}}{g_{s}}\,
\rho_I^{2}\,U_I\,\left(
\frac{
\sigma_{\mu ^{\prime}}\,\overline{x}-x_{\mu ^{\prime}}}{2\,\pi^{2}\,x^{4}}
\right)\,U_I^{\dagger}
\ 
\frac{1}{\Pi_{x}}\, ,
\label{igauge}
\\[2.4ex]
\Pi_x &\equiv & 1+\frac{\rho_I^2}{x^2} \, ,
\end{eqnarray}
depending on the various collective coordinates, the instanton size $\rho_I$
and the colour orientation matrices $U_I$. The $U_I$ matrices  
involve both colour and spinor indices, the former ranging  as usual only 
in the $2\times 2$ upper left corner of $3\times 3$ SU(3) colour matrices. 
   
\bigskip
\noindent
ii) The quark zero modes~\cite{th}, $\kappa$ and $\overline{\phi}$,
\begin{eqnarray}
\kappa^m_{\ \dot\beta}\,(x)
&=& 2\,\pi\,\rho_I^{3/2}\, \epsilon^{\gamma\delta}\,
\left( U_I\right)^m_{\ \delta}\,
\frac{\overline{x}_{\dot\beta\gamma}}{2\,\pi^2\,x^4}\ 
\frac{1}{\Pi_x^{3/2}}\, ,
\label{kappa}
\\[2.4ex]
\overline{\phi}^{\dot\alpha}_{\ l}\,(x)
&=&2\,\pi\,\rho_I^{3/2}\, \epsilon_{\gamma\delta}\,
\left( U_I^\dagger\right)^\gamma_{\ l}\,
\frac{x^{\delta\dot\alpha}}{2\,\pi^2\,x^4}\ 
\frac{1}{\Pi_x^{3/2}}\, ,
\label{phibar}
\end{eqnarray}
and

\bigskip
\noindent
iii) 
the quark propagators in the instanton background~\cite{bccl},
\begin{eqnarray}
\label{si}
\lefteqn{
S^{(I)}(x,y) =}
\\[1.6ex]
\nonumber
&&
\frac{1}{\sqrt{\Pi_x\Pi_y}}
\left[ 
\frac{x-y}{2\pi^2(x-y)^4}
\left( 1 +\rho_I^2\frac{
\left[ U_I x\,\overline{y}\,U_I^\dagger\right]}
{x^2 y^2}\right)
+\frac{\rho_I^2\,\sigma_\mu }{4\pi^2}
\frac{\left[
U_I x\,(\overline{x}-\overline{y})\,\sigma_\mu \,\overline{y}\,U_I^\dagger
\right]}
{x^2(x-y)^2 y^4\Pi_y}
\right],
\\[2.4ex]
\label{sibar}
\lefteqn{
\overline{S}^{(I)}(x,y) =}
\\[1.6ex]
\nonumber
&&
\frac{1}{\sqrt{\Pi_x\Pi_y}}
\left[ 
\frac{\overline{x}-\overline{y}}{2\pi^2(x-y)^4}
\left(  1 +\rho_I^2\frac{
\left[ U_I x\,\overline{y}\,U_I^\dagger\right]}
{x^2 y^2}\right)
+\frac{\rho_I^2\,
\overline{\sigma}_\mu }{4\pi^2}
\frac{\left[
U_I x\,\overline{\sigma}_\mu \,(x-y)\,\overline{y}\,
U_I^\dagger \right]}
{\Pi_x x^4(x-y)^2 y^2  }
\right] .
\end{eqnarray}

\begin{figure}
\begin{center}
\epsfig{file=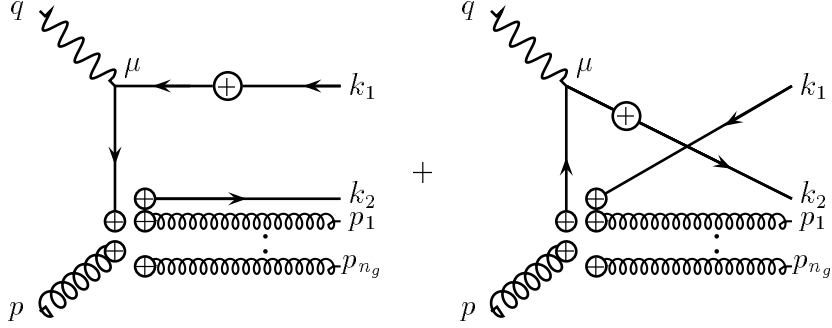}
\caption[dum]{\label{famplld}
  Instanton-induced {\it chirality-violating} process, $\gamma^{\ast}
  +{\rm g}\rightarrow 
  \overline{{\rm q}_{L}} +{\rm q}_{R} +n_{g}\,{\rm g}$,  
  for one massless
  flavour ($n_f=1$), 
  in leading semi-classical approximation. The corresponding
  Green's function involves the products of the appropriate classical fields
  (lines ending at blobs) as well as the quark propagator in the instanton
  background (quark line with central blob).}
\end{center}
\end{figure}

For simplicity, we concentrate here on the case of one flavour ($n_f=1$ in 
Eq.~(\ref{instproc})), with the generalization to a larger number of 
flavours being straightforward.

The relevant diagrams for the exclusive process of interest,
Eq.~(\ref{instproc}), are displayed in Fig.~\ref{famplld}, in leading 
semi-classical approximation.
The amplitude is expressed in terms of an
integral over the collective coordinates $\rho_I$ and
the colour orientation $U_I$,
\begin{eqnarray}
\label{ampl_inst}
\lefteqn{
  {\mathcal T}_\mu ^{a\,a_1\ldots a_{n_g}} 
  \left( \gamma^\ast +  {\rm g}\rightarrow 
    \overline{{\rm q}_L} + {\rm q}_R  +n_g\,{\rm g}\right) =} 
\\[1.6ex]
\nonumber
&&\int dU_I\,\int\limits_0^\infty
  \frac{d\rho_I}{\rho_I^5}\,d(\rho_I ,\mu _{r} )\,
  {\mathcal A}_\mu ^{a\, a_1\ldots a_{n_g}}(\rho_I, U_I)
\ ;\hspace{3.2ex} a_{(i)}=1,2,3, 
\end{eqnarray}
where
\begin{equation}
  d(\rho_I ,\mu _{r} )= d\,\left( \frac{2\,\pi}{\alpha_s(\mu _{r} )}
                        \right)^6\, 
     {\exp}\left[-\frac{2\,\pi}{\alpha_s(\mu _{r} )}\right] \,
     \left(
    \rho_I\,\mu _{r} \right)^{\beta_0}\, ,
\label{density}
\end{equation}
denotes the instanton density~\cite{th,ber,abc}, with $\mu _{r}$ 
being the renormalization scale. The density  (\ref{density}), 
with the leading-order expression for $\alpha_{s}(\mu _{r})$, 
\begin{eqnarray}
\label{alpha}
\alpha_{s}(\mu _{r}) = 
\frac{4\,\pi}{\beta_{0}\,\ln \left(\frac{\mu _{r}^{2}}{\Lambda^{2}}\right)}
\ ; \hspace{4ex}
  \beta_0 = 11-\frac{2}{3}\,n_f\, ,
\end{eqnarray}
satisfies renormalization-group invariance at the 1-loop
level\footnote{\label{2-loop} Two-loop renormalization-group invariance of 
the density (\ref{density}) can be achieved~\cite{morretal} by 
replacing in Eq.~(\ref{density}) $\beta_0$ by 
$\beta_0+\frac{\alpha_s}{4\pi}(\beta_1-12\beta_0)$, 
with next-to-leading order expression for $\alpha_s 
(\mu _r)$.}. 
The constant $d$ is given by
\begin{equation}
\label{d}
  d=\frac{C_1}{2}\,{\rm e}^{-3\,C_2+n_f\,C_3}\, ,
\end{equation}
with $C_{1}=0.466$, $C_{2}=1.54$, and $C_{3}=0.153$, in the $\overline{\rm
  MS}$-scheme. 

Before analytic continuation, the amplitude ${\mathcal A}_\mu $
entering Eq.~(\ref{ampl_inst}) takes the following form in Euclidean space,
\begin{eqnarray}
\label{ampi}
\lefteqn{
   {\mathcal A}^{a\,a_1\ldots a_{n_g}}_\mu  =
  -\ii\, e_{q} \lim_{p^{2}\rightarrow 0}p^{2}\,
    {\rm tr}\left[ \sigma^{a}
    \epsilon_g(p)\cdot A^{(I)}(p)\right]
   \prod_{i=1}^{n_g}\lim_{p_i^{2}\rightarrow 0}p_i^{2}\,
    {\rm tr}\left[ \sigma^{a_i}\epsilon_g(p_i)\cdot 
    A^{(I)}(-p_i)\right]\times }
\\[1.6ex]
  &&\chi_{R}^{\dagger}(k_{2})\, 
   \Biggl[ 
     \lim_{k_{2}^{2}\rightarrow 0}\,(\ii k_{2})\,\kappa (-k_{2})\,
     {{\mathcal V}_\mu ^{(t)}}(q,-k_{1})
 +  
    {{\mathcal V}_\mu ^{(u)}}(q,-k_{2})\,\lim_{k_{1}^{2}\rightarrow 0}\,
    \overline{\phi}(-k_{1})\,(-\ii\,\overline{k}_{1}) 
     \Biggr] 
\chi_{L}(k_{1})\, ,
\nonumber
\end{eqnarray}
with contributions ${{\mathcal V}_\mu ^{(t,u)}}$ from the diagrams on 
the left and right in Fig.~\ref{famplld}, respectively, 
\begin{eqnarray}
\label{tinst}
{{\mathcal V}_\mu ^{(t)}}(q,-k_{1})&\equiv&
\int d^{4}x\,{\rm e}^{-\ii\,q\cdot x} \,
\left[ 
\overline{\phi}(x)\,\overline{\sigma}_\mu \,
\lim_{k_{1}^{2}\rightarrow 0}\,S^{(I)}\,(x,-k_{1})\,(-\ii\,\overline{k}_{1})\,
\right] ,
\\[2.4ex]
\label{uinst}
{{\mathcal V}_\mu ^{(u)}}(q,-k_{2})&\equiv&
\int d^{4}x\,{\rm e}^{-\ii\,q\cdot x} \,
\left[
\lim_{k_{2}^{2}\rightarrow 0}\,(\ii k_{2})\,\overline{S}^{(I)}\,(-k_{2},x)\,
\sigma_\mu \,\kappa(x)\right]\, ,
\end{eqnarray}
and generic notation for various Fourier transforms involved, 
\begin{equation}
\label{fourier}
f(\ldots,k,\ldots)=\int d^4 x\,{\rm e}^{-\ii\,k\cdot x}\,
f(\ldots,x,\ldots)\, .
\end{equation}
The four-vector $\epsilon_{g\,\mu ^\prime}$ denotes the
gluon polarization-vector, whereas $\chi_{L,\,R}$ are 
Weyl-spinors, satisfying the Weyl-equations, 
\begin{equation}
\overline{k}\,\chi_L(k)=0 \, ;\hspace{0.5cm}
k\,\chi_R(k) =0 \, ,
\label{weyleq}
\end{equation}
and 
\begin{equation}
  \chi_L(k)\,\chi_L^\dagger (k) = k \, ; \hspace{0.5cm}
  \chi_R(k)\,\chi_R^\dagger (k) = \overline{k} \, .
\label{compl}   
\end{equation}

The LSZ-amputation of the classical instanton 
gauge field $A^{(I)}_{\mu ^{\prime}}$
in Eq.~(\ref{ampi}) and the quark zero modes $\kappa$ and $\overline{\phi}$
in Eqs.~(\ref{tinst}) and (\ref{uinst}), respectively, 
is straightforward~\cite{r,mrs1}.  
On the other hand, the LSZ-amputation of the quark propagators 
$S^{(I)}$ and $\overline{S}^{(I)}$ in 
Eqs.~(\ref{tinst}) and (\ref{uinst}), respectively, 
is quite non-trivial and involves a lengthy calculation. For the details, 
we refer the interested reader to Ref.~\cite{mrs1}.
Here, we only present the final result for the
scattering amplitude (\ref{ampl_inst}) in Minkowski space, 
\begin{eqnarray}
\nonumber
\lefteqn{
 {\mathcal T}_\mu ^{a\,a_1\ldots a_{n_g}}\,\left( \gamma^\ast + {\rm 
g}\rightarrow \overline{{\rm q}_L} + {\rm q}_R +n_g\,{\rm g}\right)   
 =\ii\, e_q 4 \pi^2 \left( 
\frac{\pi^3}{\alpha_s}\right)^{\frac{n_g+1}{2}} 
\int dU_I
\int\limits_0^\infty d\rho_I d(\rho_I ,\mu _{r} ) \rho_I^{2 n_g}
        } 
\\[1.6ex]
\nonumber
&&\times \,
{\rm tr}\left[ \sigma^{a}\,U_I\,\left[
\epsilon_g(p)\cdot p
-\epsilon_g(p)\,\overline{p}
\right]\,U_I^{\dagger} \right]\,
\prod_{i=1}^{n_g}
{\rm tr}\left[ \sigma^{a_i}\,U_I\,\left[
\epsilon_g(p_{i})\,\overline{p_{i}}-\epsilon_g(p_{i})\cdot p_{i}
\right]\,U_I^{\dagger} \right]
\\[1.6ex]
\label{ampi2}
&&\times 
\left\{
\left[ U_I\, \chi_R^\dagger (k_2 )\, \epsilon \right]
\left[\epsilon\,V(q,k_1;\rho_I)\overline{\sigma}_\mu \, 
\chi_L(k_1)\, U_I^\dagger
\right] \right.
\\[1.6ex]
\nonumber
&&
\left.
-
\left[ U_I\, \chi_R^\dagger (k_2)\, 
\sigma_\mu \,\overline{V}(q,k_2;\rho_I)\,\epsilon
\right]
\left[ \epsilon\, \chi_L(k_1)\, U_I^\dagger \right]
\right\} \, ,
\end{eqnarray}
where the four-vector $V_{\lambda}$ reads 
\begin{eqnarray}
  \label{Vmink}
  V_\lambda (q,k;\rho_I ) &=&
  \Biggl[  
\frac{\left( q-k\right)_\lambda}{-(q-k)^2}
+\frac{k_{\lambda}}{2 q\cdot k}
\Biggr]\rho_I\sqrt{-\left( q-k\right)^2}\,
K_1\left(\rho_I\sqrt{-\left( q-k\right)^2}\right)
\\[1.6ex]
\mbox{}&&
-
\frac{k_{\lambda}}{2 q\cdot k}\rho_I\sqrt{-q^{2}}\,
K_1\left(\rho_I\sqrt{-q^{2}}\right) \, .
\nonumber
\end{eqnarray}

At this stage of our instanton calculation, 
the constraint arising from electromagnetic gauge-invariance,
\begin{equation}
  q^\mu \,{\mathcal T}^{a\,a_1\ldots a_{n_g}}_\mu =0\, , 
\label{gaugeinv}
\end{equation}
can easily be checked. Upon contracting (\ref{ampi2}) with $q^\mu $, 
it is easy to verify Eq.~(\ref{gaugeinv}) with the help of  
the Weyl-equations (\ref{weyleq}), 
the relations
\begin{equation}
  \sigma_\mu \overline{\sigma}_\nu +\sigma_\nu \overline{\sigma}_\mu =
  2\,g_{\mu \nu} \, ,
\label{com}
\end{equation}
of the $\sigma$-matrices$^{\ref{footsigma}}$ in Minkowski-space, and the 
on-shell conditions $k_1^2=k_2^2=0$. 
 
Let us concentrate now on the main objective of this Section:
The integration over the instanton size $\rho_I$ in Eq.~(\ref{ampi2})
is {\it finite}~\cite{mrs1}. In particular, the {\it good infrared} behaviour 
(large $\rho_I$) of the integrand is due to the exponential decrease 
of the Bessel-K function for large $\rho_I$ in Eq.~(\ref{Vmink}),
\begin{equation}
\label{Bessel-K}  
K_1 ({\mathcal Q}\rho_{I} ) \stackrel{{\mathcal Q} \rho_{I}\rightarrow \infty}
{\rightarrow}\sqrt{\frac{\pi}{2}}\,\frac{1}{\sqrt{{\mathcal Q} \rho_{I}}}\,
\exp\left[-{\mathcal Q} \rho_{I}\right] 
\, .
\end{equation}
Thus, the typical hard scale in deep-inelastic scattering,
\begin{equation}
{\mathcal Q}\equiv {\rm min}\,\left\{
 Q\equiv\sqrt{ -q^2} \, ,\, 
 \sqrt{-(q-k_1)^2}\, , \,\sqrt{ -(q-k_2)^2}\right\}\, (\geq 0)
\, ,
\end{equation}
provides a dynamical IR cutoff for the instanton size, $\rho_{I}\lwig 
{\mathcal O}(1/{\mathcal Q})$ (at least in leading 
semi-classical approximation). Therefore, deep-inelastic 
scattering may be viewed as a distinguished process for studying 
manifestations of QCD-instantons. 
  
After the integration over the instanton size, which can even be performed
analytically~\cite{mrs1}, we find that the $I$-induced amplitude (\ref{ampi2})
is well-behaved as long as we avoid the (collinear) singularities which
arise when the internal quark virtualities, $-(q-k_1)^2$ (c.f. 
Fig.~\ref{famplld} (left)), or $-(q-k_2)^2$ (c.f. Fig.~\ref{famplld} (right)),
vanish. Thus, like in perturbative QCD, {\it fixed-angle scattering processes 
at high $Q^{2}$ and moderate multiplicity are reliably calculable in
(instanton) perturbation theory}~\cite{mrs1}. 

Note that the hard scale $\mathcal Q$ plays a very similar r{\^o}le as 
the vacuum expectation value of the Higgs field in electro-weak 
instanton-induced $(B+L)$-violation~\cite{th,r,m}. 
Another close analogy refers to the fact, that, similar to electro-weak 
instanton-induced $(B+L)$-violation~\cite{r,m}, 
also QCD-instanton induced processes in DIS are dominated by the 
multiple production of vector bosons (gluons in the case of QCD;
$W$'s and $Z$'s in the case of QFD). This can be seen directly from
Eq.~(\ref{ampi2}), since each additional gluon in the final state
gives rise to a factor of $(\pi^3/\alpha_s )^{1/2}$ in the 
amplitude. 

In analogy to electro-weak $(B+L)$-violation~\cite{m}, one 
expects~\cite{bb,rs} the sum of 
the final-state  gluon contributions to exponentiate, such that 
the total  $I$-induced $\gamma^\ast$g cross-section 
takes the form\footnote{\label{muf} The lower IR cutoff 
$\mu _{f}^{2}$ in the $Q^{\prime 2}$ integration in Eq.~(\ref{exp})
serves to regulate the (collinear) divergence mentioned above and
plays the r{\^o}le of a factorization scale.} 
(at large $Q^2$), 
\begin{eqnarray}
\label{exp}
\sigma^{{ (I)}}_{\gamma^\ast { g}}(x,Q^2) \equiv
\sum_{n_{ g}}\sigma^{{ (I)}}_{\gamma^\ast { g}\,n_{ g}}(x,Q^2)
\sim \int_x^1 dx^\prime\int\limits_{\mu _{f}^{2}}^{Q^2\frac{\xpr}{x}} 
\frac{dQ^{\prime 2}}{Q^{\prime 2}}\,\ldots\,
\frac{1}{Q^{\prime 2}}\,
\exp\left[-\frac{4\pi}{\alpha_s(Q^\prime)}\,F(x^\prime )\right]\, ,
\end{eqnarray}
where the so-called  ``holy-grail function''~\cite{m}
$F(\xpr )$ (normalized to F(1)=1) is expected to decrease towards 
smaller $\xpr$, which implies a dramatic growth of 
$\sigma^{{ (I)}}_{\gamma^\ast { g}}(x,Q^2)$ for decreasing $x$. 

In principle we could verify (\ref{exp}) by exploiting the 
exclusive amplitudes (\ref{ampi2}) further, i.e. by performing the 
collective-coordinate integrations, taking the modulus squared,
integrating over the appropriate phase space\footnote{\label{pub} 
This programme has been performed for the simplest $I$-induced process,
corresponding to $n_f=1, n_g=0$, in Ref.~\cite{mrs1}. The straightforward
generalization is in progress~\cite{mrs2}.}
and finally summing over $n_g$. We will, however, adopt a much more
powerful method in the next Section which, moreover, will allow us to go 
beyond the leading semi-classical approximation about the instanton 
(c.\,f. Sect.~\ref{subsec:sigqg}).

\section{\label{sec:struc} Structure Functions and HERA Cross-Section}

In this Section we want to concentrate on a first, preliminary estimate 
of the rate for $I$-induced events at HERA~\cite{mrs2} (subject to appropriate
kinematical cuts). Let us recall
that, in the one photon exchange approximation, the unpolarized 
inclusive
lepto-production cross-section is parametrized in terms of the familiar
structure functions $F_2$ and $F_L$,
\begin{equation}
\label{lepto-cross}
   \frac{d^2\sigma}{d\xbj\, d\ybj } =
  \frac{4\pi\alpha^2}{S\xbj^2\ybj^2} \left[ \left\{ 1-\ybj +\frac{\ybj^2}{2}
    \right\} F_2 (\xbj ,Q^2) - \frac{\ybj^2}{2} F_L (\xbj ,Q^2 ) \right] ,
\end{equation}
where $\sqrt{S}$ is the center-of-mass (c.m.) energy of the lepton-hadron 
system. The Bjorken variables are defined by
\begin{equation}
  \xbj \equiv \frac{Q^2}{2\,P\cdot q} ; \hspace{1cm} \ybj \equiv \frac{P\cdot 
    q}{P\cdot k} ,
\end{equation}
where $P\,(k)$ is the four-momentum of the incoming proton (lepton).

The $I$-induced contribution to the 
nucleon structure functions, e.g. $F_2^{(I )}(\xbj ,Q^2)$, is 
described~\cite{bb,rs,mrs2} in terms of the standard convolution
of {\it parton structure functions}, e.g. 
${\mathcal F}_{2\, g}^{(I )}(x,Q^2)$, with corresponding parton
densities, e.g. $f_g$,
\begin{eqnarray}
   F_2^{{ (I)}} (\xbj ,Q^2)  =   
   \sum_{p=q,g}\, \int\limits_{\xbj}^1 \frac{dx}{x}\, 
   f_{p}\left( \frac{\xbj}{x}\right)  \,\frac{\xbj}{x}\
   {\mathcal F}^{{ (I)}}_{ 2\,{ p}} { (x ,Q^2)}\, . 
\end{eqnarray}

Let us concentrate next on the calculation of the 
(dominating) $I$-induced 
contribution to the gluon structure functions~\cite{mrs2}. 

\begin{figure}
\begin{center}
\epsfig{file=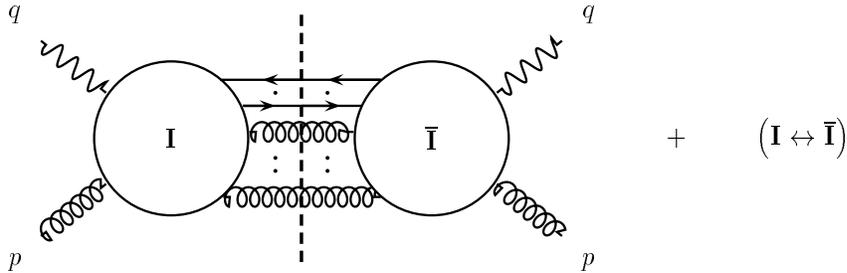,height=4.4cm}
\caption[dum]{\label{ffglgen}
Imaginary part of the  $\iai$-contribution 
to the forward $\gamma^\ast  g$ scattering amplitude.      }
\end{center}
\end{figure}

\subsection{$ I$-Contribution to the Gluon Structure Function}

The calculation rests~\cite{bb,mrs2}
on the optical
theorem combined\footnote{\label{electro}
For a combination of the optical theorem with
the $\iai$-valley method in the context
of electro-weak instanton-induced $(B+L)$-violation see 
Ref.~\cite{optvalleyqfd}.}  with the $I\overline{I}$-valley 
method~\cite{vmyung,optvalleyqcd,verb}.
The optical theorem
for the virtual $\gamma^\ast g\rightarrow \gamma^\ast g $ 
forward amplitude represents a convenient method to perform implicitly
the summation over the $I$-induced 
multi-particle final state. 

Accordingly, the $I$-induced contribution
to the gluon structure functions can be obtained by simple projections
from the imaginary part of the ${ \iai}$-contribution 
to the forward $\gamma^\ast { g}$ scattering amplitude (see 
Fig.~\ref{ffglgen}), 
\begin{eqnarray}
{\mathcal W}^{{ (I)}}_{{ g}\,\mu \nu} (p,q) &=& 
\frac{1}{\pi}\,{\rm Im}\,
{\mathcal T}^{{ (\iai)}}_{{ g}\,\mu \nu} (p,q) \,
\\[2.4ex]
\label{F2g}
{\mathcal F}^{{ (I)}}_{2\,{ g}} (x ,Q^2) &=& 
\left[ - g^{\nu\mu }\,
+ 6\,x\,\frac{p^\mu p^\nu}{p\cdot q}
\right]\,x\,
{\mathcal W}^{{ (I)}}_{{ g}\,\mu \nu} (p,q)\, ,
\\[2.4ex] 
\label{projFL}
{\mathcal F}^{(I)}_{L\,{ g}} (x ,Q^2) 
&= & 4\,x^2\,\frac{p^\mu p^\nu}{p\cdot q} 
{\mathcal W}^{(I)}_{g\,\mu \nu} (p,q) \ ,  
\end{eqnarray}
where
\begin{eqnarray}
x\equiv \frac{Q^2}{2\,p\cdot q} 
\end{eqnarray}
is the 
Bjorken variable of the  $\gamma^\ast\, g$ subprocess.

In order to extract the required ${ \iai}$-contribution 
(c.f. Fig.~\ref{ffglgen}) from the forward $\gamma^\ast g$ scattering 
amplitude, one first evaluates the path-integral expression for the 
corresponding Green's function in Euclidean space by expanding about the   
$\iai$-configuration $A_\mu ^{(\iai )}$ ($\iai$ ``valley''), which is defined 
via the so-called ``valley method\footnote{\label{footvalley} For any fixed 
values of the $\iai$ collective
coordinates $\tau_i$, $\tau =\{\rho_I, \rho_{\overline{I}},  R_\mu ,
U \}$, the pair configuration 
$A_\mu ^{(\iai)}$ is required to minimize
the action within the subspace orthogonal to 
$\partial A_\mu ^{(\iai)}/\partial
\tau_i$.}''~\cite{vmyung,optvalleyqcd,verb}.  
After Fourier transforming to momentum space, one has to amputate the 
external legs and to analytically continue
to Minkowski space. Finally, the imaginary part is taken. 

\begin{figure}
\begin{center}
\epsfig{file=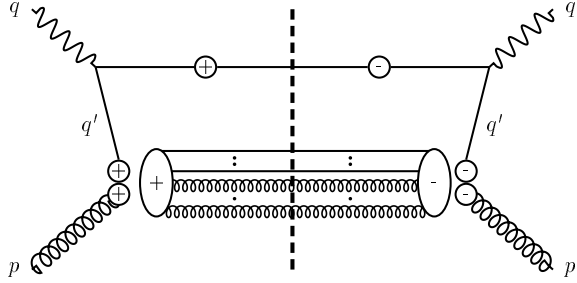,height=4.4cm}
\caption[dum]{\label{ffglld}
Structure of the 
imaginary part of the  $\iai$-contribution 
to the forward $\gamma^\ast  g$ scattering amplitude, in leading 
semi-classical approximation about the 
$\iai$-valley and at large $Q^2$.}
\end{center}
\end{figure}

Along these lines, it is possible to show~\cite{mrs2} that, in leading 
semi-classical approximation about the $\iai$-valley and at large $Q^{2}$, the 
dominant projection contributing to ${\mathcal F}_{2\,{ g}} (x ,Q^2)$
has the 
following structure (see Fig.~\ref{ffglld}),
\begin{eqnarray}
\nonumber
\lefteqn{ 
-g^{\mu \nu}
{\mathcal W}_{g\,\mu \nu}^{(I)} (p,q )         
\simeq      
\sum_q e^2_q \frac{\pi}{\alpha_s}
\int d^4 q^\prime \delta^{(+)} \left( (q-q^\prime )^2\right) 
\frac{Q^2 (p\cdot q^\prime )^2}
{\left( (p+q^\prime )^2 \right)^{3/2}} 
\left\{ 1 + {\mathcal O}\left( \frac{Q^{\prime 2}}{Q^2}\right)\right\}
  }
\\[1.6ex]
\label{formfglld}  
&&\times \,
\int dU \int\limits_{0}^{\infty} \frac{d\rho_I}{\rho_I^5} 
\int\limits_{0}^{\infty} \frac{d\rho_\ai}{\rho_\ai^5} 
\int d^4 R\, {\mathcal M}(\rho_I ,\rho_\ai , \xi, U )\, 
\exp\left[-\frac{4\pi}{\alpha_s(\mu _r )} S^{(\iai )}(\xi ,U)
     \right]
\\[1.6ex]
\nonumber
&&\times \,
\exp\left[\ii\,(p+q^\prime)\cdot R\right]\,
K_1\left( Q^{\prime }\,\rho_I\right)
       K_1\left( Q^{\prime}\,\rho_\ai\right)\, 
\left( \rho_I \rho_\ai \right)^{9/2} \,
\left( \frac{16}{\xi^3}\right)^{n_f-\frac{1}{2}}\,
\, ,
\end{eqnarray}
where $Q^{\prime 2} \equiv -q^{\prime 2}$. 
The second projection~(\ref{projFL}), entering 
${\mathcal F}_{L\,{ g}} (x ,Q^2)$, may be evaluated
analogously~\cite{mrs2}. It turns out to vanish in the Bjorken limit.
    
In analogy to the single instanton case discussed in 
Sect.~\ref{whatisspecial}, the amplitude (\ref{formfglld}) is expressed 
in terms of an integral over the appropriate $\iai$ collective coordinates, 
the $I(\ai )$-size parameters, $\rho_{I(\ai )}$, the
distance vector $R_\mu $ between $I$ and $\ai$, and the relative colour 
orientation, $U$.  
Due to conformal invariance of the classical
Yang-Mills action, the $\iai$-valley action $S^{(\iai )}(\xi , U )$
depends on the distance
$R_\mu $ and the $I$- and $\ai$-sizes only
in a conformally invariant manner~\cite{vmyung},
\begin{equation}
\label{ximink}
\xi =  {-R^2 +\ii\epsilon\,R_0 + \rho_I^2 +\rho_\ai^2\over 
\rho_I\rho_\ai }\, ; \hspace{4.0ex}({\rm in\ Minkowski\ space})\, .
\end{equation}
The $\iai$-valley action itself, for the most attractive relative
orientation, $U_0=\frac{R}{\sqrt{-R^2+\ii\,\epsilon\,R_0}}$, is given
by~\cite{optvalleyqcd,verb},
\begin{eqnarray}
\label{action}
\lefteqn{
S^{(\iai )}(\xi ,U_0)\equiv}
\\[1.6ex]
\nonumber
&& S^{(\iai )}(\xi )   =
 1-\frac{12}{f(\xi )} 
- \frac{96}{f(\xi )^2} +\frac{48}{f(\xi )^3}\left[ 3f(\xi )+8\right]
\ln\left[ \frac{1}{2\xi }\bigl( f(\xi ) +4\bigr)\right] \, ,
\end{eqnarray}
\begin{equation}
f(\xi )=\xi^2+\sqrt{\xi^2-4}\xi-4 \,  .     
\end{equation}
In Euclidean space, where $-R^2 +\ii\epsilon\,R_0$ in
(\ref{ximink}) is replaced by $R^2$, 
the $\iai$-valley action $S^{(\iai )}(\xi )$   
corresponds to a smooth interpolation 
between a widely separated,
weakly interacting $\iai$-pair configuration 
($S^{(\iai )}(\xi \rightarrow \infty)=1$) and 
a strongly overlapping one, annihilating to the perturbative 
vacuum ($S^{(\iai )}(\xi \rightarrow 2)=0$).

Finally, the combination    
\begin{eqnarray} 
\label{iaidensity} 
\lefteqn{
 {\mathcal M}(\rho_I ,\rho_\ai , \xi ,U )\,
\exp \left[-\frac{4\pi}{\alpha_s(\mu _r )} S^{(\iai )}(\xi ,U)\right]
 \simeq }
\\[1.6ex]
\nonumber
&&
d^2\, 
\biggl( {2\pi\over \alpha_s 
( \mu _r )}\biggr)^{12}\,
\exp \left[-\frac{4\pi}{\alpha_s(\mu _r )} S^{(\iai )}(\xi ,U)\right] 
\, 
\left( \rho_I \rho_\ai\, \mu _r^2\right)^{\beta_0\,S^{(\iai )}(\xi , U)}
\, ,
\end{eqnarray}
appearing in the $\iai$-induced amplitude (\ref{formfglld}), denotes the
(one-loop renormalization-group invariant) $\iai$-density\footnote{
\label{footiaidensity} There are corrections to Eq.~(\ref{iaidensity}) which
vanish for large $\xi$.  
Fortunately, we need the density only at sufficiently large $\xi$ (see below),
such that Eq.~(\ref{iaidensity}) is approximately valid.}.

After having defined the quantities entering 
Eq.~(\ref{formfglld}), let us come now to its physical interpretation. 
Complementary to the result quoted in Ref.~\cite{bb}, 
our result yields a momentum-space picture of the 
$\iai$-contribution to the imaginary part of the forward $\gamma^{\ast}g$ 
amplitude. As is clear from the discussion in 
Sect.~\ref{whatisspecial}, the
$\delta^{(+)}$-function as well as the Bessel-K functions appearing 
in Eq.~(\ref{formfglld}) are easily understood as originating from the 
discontinuity across the 
handle in the $\iai$-induced ``handbag'' diagram in Fig.~\ref{ffglld}.
The factor $\exp\left[ \ii\,(p+q^\prime)\cdot R\right]$, on the other
hand, arises from the external gluons ${\rm g}(p)$ and the internal, virtual 
quarks ${\rm q}^{\ast}(q^{\prime})$ in the $\iai$ background. 
It is then tempting to introduce the notion of an 
{\it $I$-induced $q^\ast g$ subprocess}, along with its associated 
total cross-section~\cite{mrs2},
\begin{eqnarray}
\label{qgcross}
\lefteqn{  
\sigma_{q^\ast g}^{(I )}(\xpr ,Q^{\prime 2})         
\equiv     
\frac{8}{3}\,\frac{\pi^5}{\alpha_s}\,
\frac{Q^{\prime 2}\,(p\cdot q^\prime )^2}
{\left( (p+q^\prime )^2 \right)^{3/2}}\,
  }
\\[1.6ex]
\nonumber
&&\times \,
\int dU \int\limits_{0}^{\infty} \frac{d\rho_I}{\rho_I^5} 
\int\limits_{0}^{\infty} \frac{d\rho_\ai}{\rho_\ai^5} 
\int d^4 R\, {\mathcal M}(\rho_I ,\rho_\ai , \xi ,U )\, 
\exp\left[ -\frac{4\pi}{\alpha_s(\mu _r )} S^{(\iai )}(\xi ,U)
     \right]
\\[1.6ex]
\nonumber
&&\times \,
\exp\left[ \ii\,(p+q^\prime)\cdot R\right]\,
K_1\left( \sqrt{Q^{\prime 2}}\,\rho_I\right)
       K_1\left( \sqrt{Q^{\prime 2}}\,\rho_\ai\right)\, 
\left( \rho_I \rho_\ai \right)^{9/2} \,
\left( \frac{16}{\xi^3}\right)^{n_f-\frac{1}{2}}\,
\, .
\end{eqnarray}
Upon partially performing the integration over $q^\prime_\mu $ in 
Eq.~(\ref{formfglld}), one may easily show that, 
for large $Q^{2}$, 
the $I$-contribution to the gluon structure function ${\mathcal F}_{2\, g}$, 
Eq.~(\ref{F2g}), can be expressed as~\cite{mrs2}
\begin{equation}
{\mathcal F}_{2\, g}^{(I )} (x ,Q^2) \simeq \sum_{ q} e_{ 
q}^2\ x\, \int\limits_x^1 \frac{dx^\prime}{x^\prime}
\,\left( \frac{3}{8\,\pi^3}\,\frac{x}{x^\prime}\right)\, 
\int\limits_{\mu _f^2}^{Q^2\frac{x^\prime}{x}}
dQ^{\prime 2} \, 
 \sigma_{q^\ast  g}^{(I )}
(x^\prime, Q^{\prime 2}) \, , 
\label{convolution}
\end{equation}
where
\begin{equation}
\xpr = \frac{Q^{\prime 2}}{2\,p\cdot q^\prime } 
\end{equation}
denotes the Bjorken variable of the $q^{\ast}g$ subprocess.
Thus, in leading-order semi-classical approximation about the 
$\iai$-valley and at large $Q^2$, the $I$-contribution
to the gluon structure function ${\mathcal F}_{2\, g}$ has the form of a 
convolution of a ``splitting function'' in the $I$-background,
\begin{equation}
P_{q^\ast/\gamma^\ast}^{(I)}\,\left( 
\frac{x}{x^\prime}\right) \equiv \frac{3}{8\,\pi^3}\,\frac{x}{x^\prime}\, ,
\end{equation} 
with a total cross-section (\ref{qgcross}) for the $I$-induced subprocess,  
containing the essential instanton dynamics.  
A generalization of this convolution structure~\cite{rs} to less 
inclusive quantities, e.g. single-particle inclusive distributions, is 
in progress~\cite{mrs2}.

\subsection{\label{subsec:sigqg} 
             The Total Cross-Section of the $I$-Induced $q^\ast g$ Subprocess}

In this Section we shall concentrate on the most important building
block entering  Eq.~(\ref{convolution}), the total
cross-section (\ref{qgcross}) of the $I$-induced  $q^\ast g$ subprocess 
(c.\,f. Fig.~\ref{ffglld}). 

The collective coordinate integrals in Eq.~(\ref{qgcross}) can be 
done in the saddle-point approximation, the small parameter being 
$\alpha_s (\mu _r )$. The expression to be extremized is
the exponent in Eq.~(\ref{qgcross}), 
\begin{equation}
\label{gamma}
\Gamma (p,q^\prime ;\rho_I,\rho_\ai, R_\mu , U) \equiv
\ii\,(p_g+q^\prime)\cdot R  
-\,Q^\prime \left(\rho_I+\rho_\ai\right)
-\frac{4\pi}{\alpha_s\left(\mu _r \right)} S^{(\iai )}(\xi , U)
\, ,
\end{equation}
where we have used the asymptotic form (\ref{Bessel-K}) for the Bessel-K 
functions, anticipating that,   
for small $\alpha_s (\mu _r )$, the dominant contribution
to Eq.~(\ref{qgcross}) will come from the region
\begin{equation} 
Q^\prime\, \rho_{I(\ai )}\gg 1\, .
\label{Qrholarge}
\end{equation}

The corresponding saddle-point equations are most 
easily solved in the $q^\ast g$ c.\,m. system, where the solution
is given by~\cite{optvalleyqcd,bbgg},
\begin{eqnarray}
\label{xiast}
\xi^\ast &=& 2 +  \frac{4}{\omega^{\prime 2}} \, ,
\\[2.4ex]
\label{rhoast}
\rho^\ast_I = \rho^\ast_\ai &=& 
\frac{16\,\pi}{\alpha_s (\mu _r )}\,
\frac{\frac{dS^{(\iai )}}{d\xi}(\xi^\ast )}{\omega^{\prime 2}}\, 
\frac{1}{Q^\prime} \, ,
\\[2.4ex]
\label{Rmuast}
R_\mu ^\ast &=& \left( -\ii\,
\sqrt{{\rho}_I^\ast{\rho}_\ai^\ast}\,
\sqrt{\xi^\ast -2-\frac{({\rho}_I^\ast -{\rho}_\ai^\ast)^2}
{{\rho}_I^\ast{\rho}_\ai^\ast}}
\, ,\, \vec{0} \right)
\, ,
\\[2.4ex]
\label{Uast}
U^\ast &=& \frac{R^\ast}{\sqrt{-R^{\ast\,2}}} \, .
\end{eqnarray}
Here $\omega^{\prime}$ denotes a scaling variable corresponding to the
$q^\ast g$ c.\,m. energy, $\sqrt{s^\prime}$,
\begin{equation}
\label{omegaprime}
\omega^{\prime} \equiv \frac{\sqrt{s^\prime}}{Q^\prime} = 
\sqrt{\frac{1-\xpr}{\xpr}}
\, .
\end{equation}
Note, that the saddle-point for $\rho$, Eq.~(\ref{rhoast}), is consistent
with our ansatz (\ref{Qrholarge}), as long as $\alpha_s (\mu _r )$ is 
small.

\begin{figure}
\begin{center}
\epsfig{file=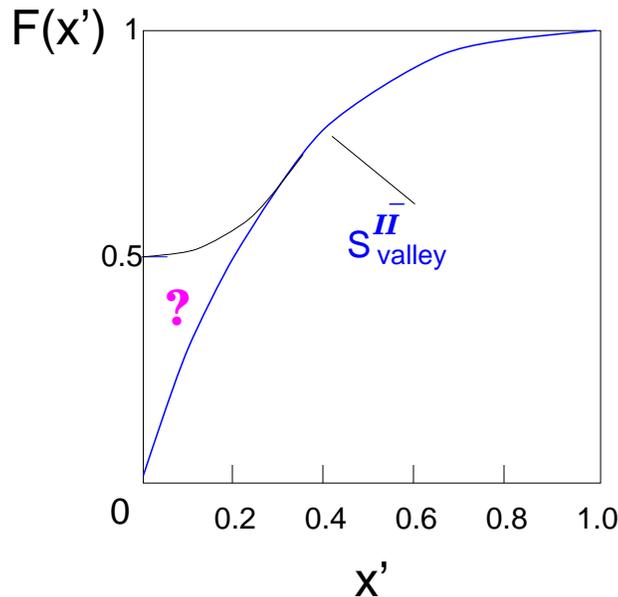,height=8cm}
\caption[dum]{\label{fsvalley}
The holy-grail function and the valley action.}
\end{center}
\end{figure}

The exponent $\Gamma$, Eq.~(\ref{gamma}), evaluated at the saddle-point,
Eqs.~(\ref{xiast}-\ref{Uast}), takes a particularly 
simple form~\cite{optvalleyqcd,bbgg}, 
\begin{equation}
\label{gammaast}
\Gamma (p,q^\prime ;\rho_I^\ast ,\rho_\ai^\ast, 
             R^\ast_\mu , U^\ast ) =
-\frac{4\pi}{\alpha_s\left(\mu _r \right)} S^{(\iai )}\left(\xi^\ast 
(\xpr )\right) \, .
\end{equation}
It depends only on the scaling variable $\xpr$ 
(or, equivalently, on $\omega^{\prime}$).  

Now we are ready to perform the integrations about the
saddle-point. The final result for the cross-section (\ref{qgcross}), 
for small $\alpha_s (\mu _r )$, reads~\cite{mrs2},
\begin{equation}
\sigma_{q^\ast g}^{(I )} (\xpr ,Q^{\prime 2})
\simeq \frac{\Sigma (\xpr )}{Q^{\prime 2}}\ 
\left( \frac{4\pi}{\alpha_s (\mu _r )}\right)^{21/2+2\,\beta_{0}\,F(\xpr )}
\ \exp\left[-\frac{4\pi}{\alpha_s (Q^{\prime} )}
\,F(\xpr )\right]
\, ,
\label{sigma}
\end{equation}
where the ``holy-grail" function $F(\xpr)$ is here identified as 
(see Eq.~(\ref{gammaast}) and Fig.~\ref{fsvalley}),
\begin{equation}
\label{holygrail}
F(\xpr ) \equiv S^{(\iai )} \left(\xi^\ast (\xpr )\right)
\, .
\end{equation}
Moreover, we find~\cite{mrs2},
\begin{eqnarray}
\Sigma (\xpr ) &= &
 d^2\,\sqrt{6}\ 2^{n_f-4}\,\pi^{15/2}
\frac{\xpr^{11/2}}{(1-\xpr)^7}\,
\left( \frac{1-\xpr}{1+\xpr} \right)^{3n_f+2} 
\left[
\xpr\,
(1-\xpr)\,\frac{dF(\xpr )}{d\xpr}\right]^{2\,\beta_0\,F(\xpr )-5/2} \, .
\label{prefacxprime} 
\end{eqnarray}

\begin{figure}
\begin{center}
\epsfig{file=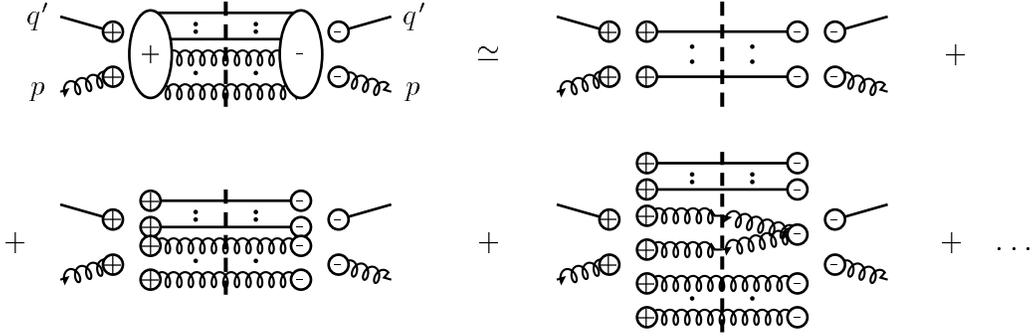}
\caption[dum]{\label{fhgpert}
Low-energy expansion of the total instanton-induced subprocess
cross-section. The three different graphs on the rhs. correspond
to the respective three different terms in the low-energy expansion
of the holy-grail function (\ref{holypert}). 
}
\end{center}
\end{figure}

Let us investigate our result for the holy-grail function $F(\xpr )$,
Eq.~(\ref{holygrail}),
for small $q^\ast g$ c.m. energy $\sqrt{s^\prime}$, i.e. small 
$\omega^{\prime}$ (c.f. Eq.~(\ref{omegaprime})), 
\begin{eqnarray}
\label{holypert}
F(\xpr )=1-\frac{3}{8}\left( \frac{1-\xpr}{\xpr}\right)^2
+\frac{3}{8}\left( \frac{1-\xpr}{\xpr}\right)^3+{\mathcal O}
\left( \left( \frac{1-\xpr}{\xpr}\right)^4\right) 
.
\end{eqnarray}
The origin of the different terms on the rhs. of Eq.~(\ref{holypert}) 
is illustrated in Fig.~\ref{fhgpert}.
The first term (`t Hooft tunneling factor~\cite{th}) corresponds to a 
purely fermionic final state. 
The next term arises from the summation of the 
leading-order gluon emission (c.f. Section~\ref{whatisspecial}), and the
third term originates~\cite{holypert} from the summation of interference 
terms between the leading-order gluon emission and the gluon-propagator 
correction, see Fig.~\ref{fhgpert}.
 
The $\iai$-valley method allows to extrapolate (\ref{holypert}) 
smoothly beyond instanton-perturbation theory, via the identification of 
the holy-grail function with the $\iai$-valley action at the saddle point, 
Eq.~(\ref{holygrail}). As we have illustrated in Fig.~\ref{fhgpert}, it 
effectively sums up the gluonic final-state tree-graph 
corrections to the leading semi-classical result\footnote{\label{hard} For 
a formal proof, see first reference in Ref.~\cite{holypert}.}. 
However, it has been argued~\cite{hard} that some 
initial-state and initial-state - final-state corrections exponentiate as well
and might give rise to additional corrections, which show up in the
low-energy expansion of the holy-grail function (\ref{holypert}) at order
${\mathcal O}(\omega^{\prime 5})$.
Nevertheless, it appears reasonable to trust our result 
(\ref{sigma}-\ref{prefacxprime}) down to $\xpr= 0.2$,  
where $F(0.2)\equiv S^{(\iai )}(\xi^{\ast}(0.2))\simeq 1/2$ 
(see Fig.~\ref{fsvalley}),  
a value sometimes advocated \cite{hgbound}
as the lower bound for the holy-grail function.

\begin{figure}
\begin{center}
\epsfig{file=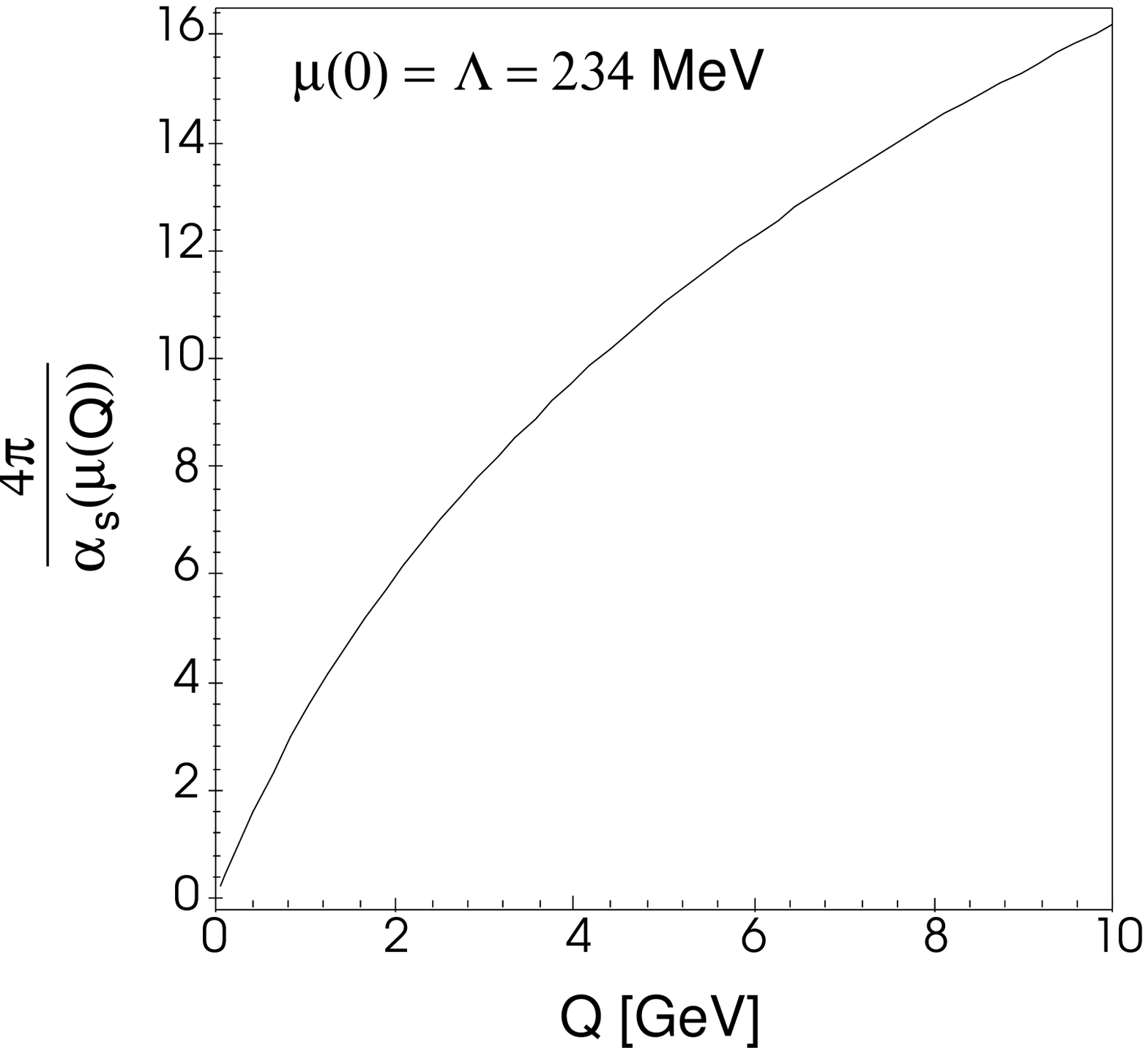,width=7.6cm}\hfill
\epsfig{file=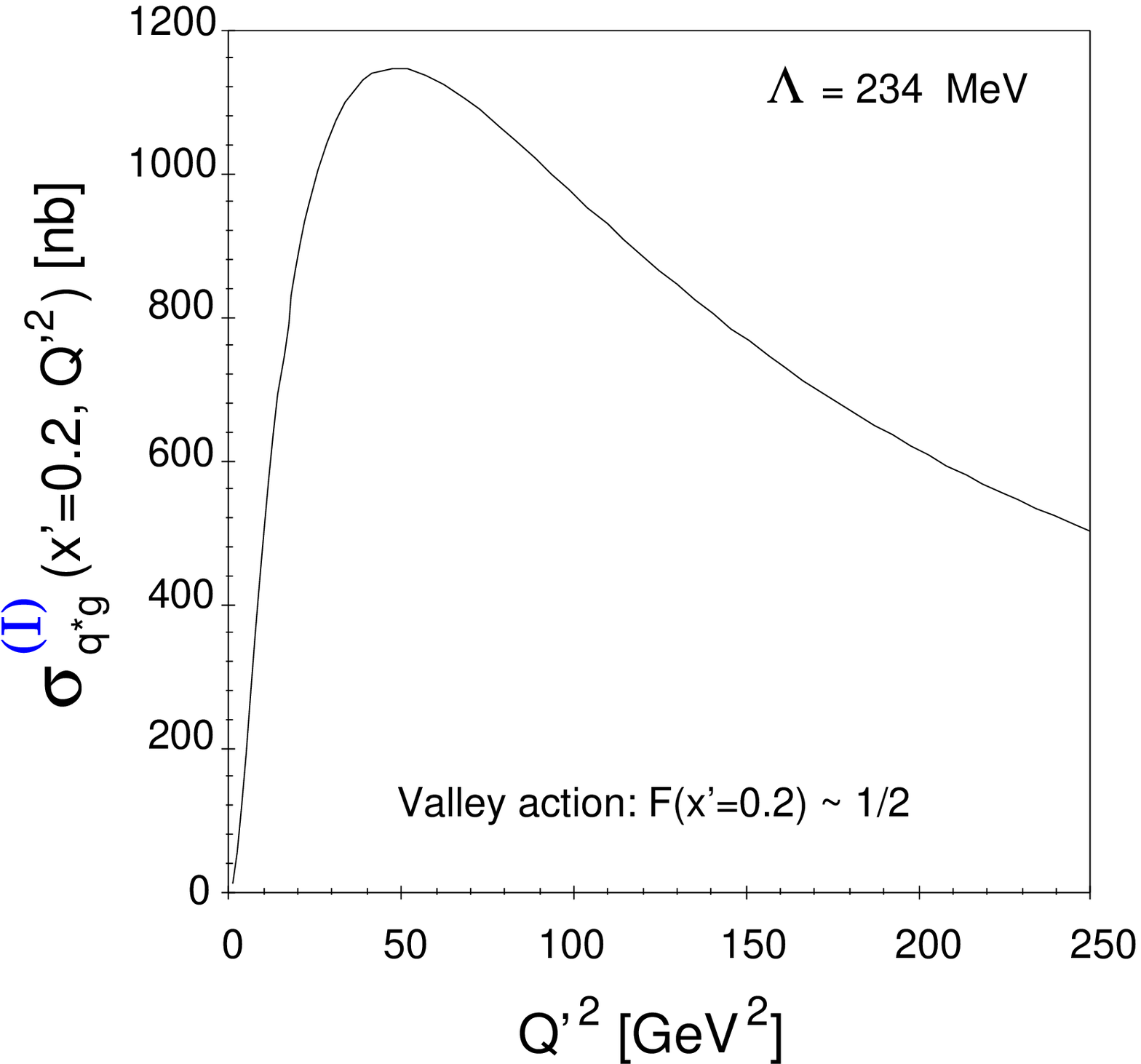,width=7.6cm}
\caption[dum]{\label{falpha} 
Left: The inverse running coupling at the renormalization scale $\mu (Q)$, 
acting as the large parameter in the saddle-point approximation. 
Right: The total cross-section of the $I$-induced $q^\ast g$ subprocess, 
$\sigma_{q^\ast g}^{(I )}(\xpr ,Q^{\prime 2})$, for fixed $\xpr =0.2$. 
 }
\end{center}
\end{figure}

As it stands, our result (\ref{sigma}) for the $I$-induced total
$q^{\ast} g$ cross-section strongly depends on the renormalization
scale $\mu _{r}$. Possible improvements, partly along the lines indicated in
footnote~\ref{2-loop}, will be discussed elsewhere~\cite{mrs2}. 
The main $\mu _{r}$-dependence in Eq.~(\ref{sigma}),
\begin{equation}
\left[ \frac{4\pi}{\alpha_s (\mu _r )}\right]^{2\,\beta_{0}\,F(\xpr )}
=
\left[ \beta_{0}\, 
\ln \left( \frac{\mu _{r}^{2}}{\Lambda^{2}}\right)
\right]^{2\,\beta_{0}\,F(\xpr )}
\, ,
\end{equation}
has its origin in the instanton-size dependence of the
$\iai$-density (\ref{iaidensity}), 
$(\rho_{I}\rho_{\ai}\mu _{r}^{2})^{\beta_{0}S^{(\iai)}}$, taken at the saddle
point (\ref{rhoast}). A natural choice~\cite{bb,bbgg} for the 
renormalization scale is $\mu _{r}=1/\rho^{\ast}$. A closely related choice, 
which we shall adopt in the following, is to fix the renormalization scale,
\begin{equation}
\label{murchoice}
\mu _{r}=\mu (Q^{\prime})\, ,
\end{equation}
by the requirement,
\begin{equation}
\label{muQprime}
\frac{4\,\pi}{\alpha_{s}(\mu (Q^{\prime}))}\,
\frac{\mu (Q^{\prime})}{Q^{\prime}} =
1 
\, .
\end{equation}
With this choice, the cross-section 
reads~\cite{rs1,mrs2},
\begin{eqnarray}
\sigma_{q^\ast g}^{(I )} (\xpr ,Q^{\prime 2})
&\simeq& \frac{\Sigma (\xpr )}{Q^{\prime 2}}\ 
\left( \frac{4\pi}{\alpha_s (\mu (Q^{\prime}) )}\right)^{21/2}
\ \exp\left[-\frac{4\pi}{\alpha_s (\mu (Q^{\prime}) )}
\,F(\xpr )\right]
\, ,
\label{sigmala}
\\[15pt]
\nonumber
&&\mbox{for}\ \xpr\gwig\, 0.2 \ ,\ Q^{\prime}\gwig\, 5\ {\rm GeV}
\ .
\end{eqnarray}
The restriction on $\xpr$ results mainly from our insufficient knowledge of
the holy-grail function $F(\xpr )$ for small $\xpr$, whereas the
restriction on $Q^{\prime}$ is due to the requirement that the 
parameter $4\pi/\alpha_{s}(\mu (Q^{\prime}))$ should be large 
(see Fig.~\ref{falpha} (left)) in order to justify the saddle-point 
approximation.

Note the following important feature of 
$\sigma_{{ q}^\ast { g}}^{(I )}(\xpr ,Q^{\prime 2})$, 
Eq.~(\ref{sigmala}), as a function of 
$Q^{\prime}$ (see Fig.~\ref{falpha} (right)): The $Q^{\prime}$ dependences 
from the high inverse power\footnote{\label{power} The large power $21/2$ 
mainly reflects the number $12$ of $\iai$ zero modes minus $1/2$ times 
the number of saddle-point integrations.} 
of $\alpha_{s}$ and the exponential in Eq.~(\ref{sigmala})
compete to produce a strong peak (maximum) far away from the IR region, 
at a new {\it hard} scale, 
\begin{equation}
\Lambda^{(I)} (\xpr ) \equiv 2\,\beta_{0}\,
\frac{\frac{21}{4} -1 }{1+\beta_{0}\,F(\xpr )}\,
\exp\left\{ 
\frac{ \frac{21}{4}-1}{1+\beta_{0}\,F(\xpr )}\right\}\,
\Lambda\ \gg \Lambda\ ,
\label{newscale}
\end{equation}
much above the usual QCD scale $\Lambda$. This implies that
the $I$-contribution to the gluon structure function 
${\mathcal F}_{2\,{ g}}^{(I )}(x ,Q^2)$, which involves an integration of 
$\sigma_{{ q}^\ast { g}}^{(I )} (\xpr ,Q^{\prime 2})$ over $Q^{\prime 2}$ 
(c.f. Eq.~(\ref{convolution})), 
\begin{eqnarray}
\label{Qprimeint}
\lefteqn{
{\mathcal F}_{2\,{ g}}^{(I )} (x ,Q^2) \simeq 
\sum_{ q} e_{ q}^2\ x\,
\int_x^1 \frac{dx^\prime}{x^\prime}\,
P^{(I )}_{{ q}^\ast/\gamma^{\ast}}\left( \frac{x}{x^\prime} \right)\,
\Sigma (\xpr ) } 
\\[1.6ex] 
\nonumber
&&\times \, 
\int\limits_{\mu _{f}^{2}}^{Q^2\frac{\xpr}{x}} 
\frac{dQ^{\prime 2}}{\Qprime^2}\,
\left( \frac{4\pi}{\alpha_s (\mu (Q^\prime ) )}\right)^{21/2}
\ \exp\left[-\frac{4\pi}{\alpha_s (\mu (Q^\prime ) )}
\,F(\xpr )\right]
\, ,
\end{eqnarray}
is asymptotically dominated by this peak and hence $Q$ independent (scaling)
in the Bjorken limit\footnote{Note, that the simplest $I$-induced (exclusive) 
process, without gluons in the final state (c.\,f. Sect.~\ref{whatisspecial}), 
already gives a scaling contribution~\cite{mrs1} to the gluon structure 
function ${\mathcal F}_{2\,{ g}}^{(I )}(x ,Q^2)$.},
\begin{eqnarray}
{\mathcal F}_{2\,{ g}}^{(I )} (x ,Q^2)  
\stackrel{Q\ {\rm large}}{\approx }
\sum_{ q} e_{ q}^2\ x\,
\int_x^1 \frac{dx^\prime}{x^\prime}\,
P^{(I )}_{{ q}^\ast/\gamma^{\ast}}\left( \frac{x}{x^\prime} \right)\,
\Sigma (\xpr )\ 2\ \Gamma \left(\frac{21}{2}\right)\,
\frac{F(\xpr )+\frac{21}{4\,\beta_{0}}}{F(\xpr )^{23/2}} 
\label{Qprimeintasy}
 \, .
\end{eqnarray}
The predicted approach to scaling,
\begin{eqnarray}
\lefteqn{
Q^2\ \frac{d\,
{\mathcal F}_{2\,{ g}}^{(I )} (x ,Q^2) }{dQ^2} 
\stackrel{Q\ {\rm large}}{\approx } }
\\[1.6ex]
\nonumber
&&
\sum_{ q} e_{ q}^2\, x
P^{(I)}_{{ q}^\ast/\gamma^{\ast}}\left( 1 \right)
\frac{\Sigma (x )}  
{ \frac{dF(x)}{dx} } 
\left[ \beta_{0}\,\ln\left( \frac{Q^2}
             {\Lambda^2}\right)\right ]^{\frac{19}{2}+2\,\beta_{0}\,F(x)}
             \left [\frac{\Lambda^2}{Q^2}\right ]^{\beta_{0}\,F(x )}
\, , 
\end{eqnarray}
resembles a ``fractional twist'' term, where 
the twist is sliding with $x$: the scaling violations vanish faster  with 
increasing $x$. 

Note, however, that the above results for the gluon structure function 
depend to some extent  on our choice of
the renormalization scale $\mu _{r}=\mu (Q^{\prime })$, 
Eqs.~(\ref{murchoice}), (\ref{muQprime}). A thorough investigation
of the inherent renormalization- and factorization-scale dependences is
presently under way~\cite{mrs2}. Then, we may also compare in more detail with
the results of Ref.~\cite{bb} and sort out possible differences. 

\begin{figure}
\begin{center}
\epsfig{file=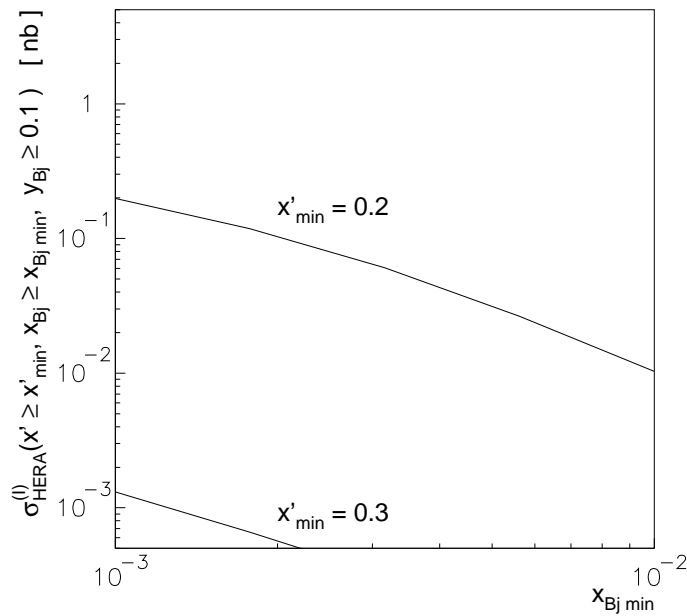,%
bbllx=15pt,bblly=149pt,bburx=537pt,bbury=647pt,height=8cm,angle=0}
\caption[dum]{\label{fsigmacut}
Instanton-induced total $e p$ cross-section for HERA (preliminary) with 
various cuts as indicated. 
}
\end{center}
\end{figure}

\subsection{HERA Cross-Section}

In Fig.~\ref{fsigmacut} we present the resulting $I$-induced 
total $e p$ cross-section for HERA, {\it subject to the following cuts:}
\begin{itemize}
\item $\xpr \ge \xpr_{\rm min}= (0.2,0.3)$ 
(c.f. discussion in Sec.~\ref{subsec:sigqg})
in the integration of Eq.~(\ref{Qprimeint});
\item $x_{\rm Bj}\ge x_{\rm Bj\ min}$, $y_{\rm Bj}\ge 0.1$ in the integration 
of Eq.~(\ref{lepto-cross}). 
\end{itemize}
So far, only the (dominating) gluon contribution has been taken into account.

Apparently, the cross-section is surprisingly
large. Let us, however, emphasize again that
the inherent uncertainties associated with the renormalization- and 
factorization-scale dependences may be considerable~\cite{mrs2}.
Therefore, Fig.~\ref{fsigmacut} is still to be considered preliminary.  
Furthermore, there is, of course, also a strong dependence on the precise value
of the QCD scale $\Lambda$ (here taken to be 234 MeV).

\begin{figure}
\begin{center}
\epsfig{file=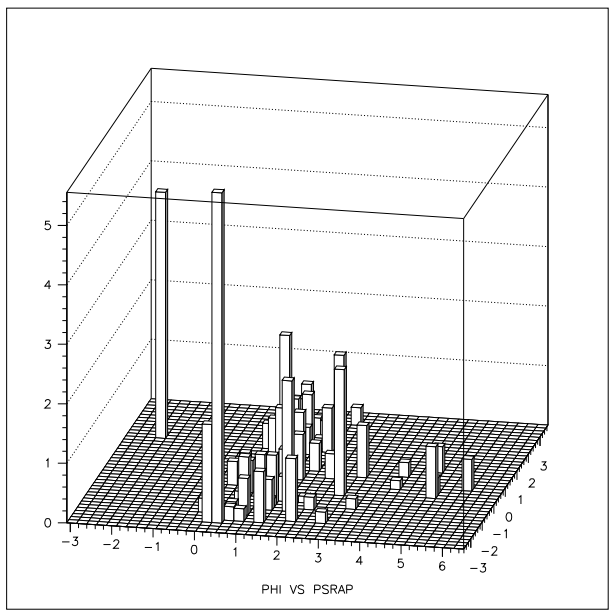,height=8cm}
\caption[dum]{\label{flego}
Lego plot $(\eta_{\rm lab},\phi_{\rm lab},E_{T}[{\rm GeV}])$ 
of a typical $I$-induced event in the HERA-lab system at $x_{\rm Bj}=10^{-3}$.
 }
\end{center}
\end{figure}

\section{\label{sec:monte} Monte-Carlo Simulation for HERA}

A Monte-Carlo generator for QCD-instanton induced events in DIS, QCDINS~1.3 
based on HERWIG~5.8, has been developed\footnote{
Until the final write-up in Ref.~\cite{grs2} is completed, Ref.~\cite{ewmc}
may be consulted as a qualitative guide for QCDINS. The Monte-Carlo event
generator for electro-weak instanton-induced processes (HERBVI) 
in Ref.~\cite{ewmc} shares a number of similarities with QCDINS.}
~\cite{grs,grs2}.
It has the following features built-in, which are characteristic for the 
underlying instanton mechanism:

\begin{itemize}
\item In its c.m. system, $\vec{q^{\prime}}+\vec{p}=0$, the 
      $I$-induced multi-parton production is supposed to
      proceed {\it isotropically}~\cite{th,r,m,bb,rs1}. 
      We may imagine a ``fireball''
      in $S$-wave configuration, decaying into gluons and at
      least $2\,n_{f}-1$ quarks.
\item The total parton multiplicity associated with the $I$-subprocess
      is large~\cite{r,m,bb,rs1,mrs2}, 
       \begin{equation}
           \langle n_{q+g}\rangle^{(I)} ={\mathcal O}(10)\ \ {\rm at\ HERA}.
       \end{equation}
\item The multiplicity distribution at the parton level is taken to be  
      a Poisson distribution\footnote{Strictly speaking, this holds only in
      the Bjorken limit~\cite{mrs2}.}~\cite{mrs2}.
\item All light flavours, including strangeness and possibly 
charm\footnote{\label{charm}
The predictions associated with the $I$-induced production of charmed
quarks are less certain, since for $m_q \rho^\ast\gwig 1$ additional
suppression factors may arise.}, if kinematically allowed, are 
democratically produced~\cite{th}.
\end{itemize}       
At present, we consider the characterization of $I$-induced events by 
these final-state features to be more robust than predictions based on
cross-section estimates.

The typical event in Fig.~\ref{flego} 
from our Monte-Carlo generator~\cite{grs2} 
illustrates these built-in characteristics:
A current-quark jet along with a densely populated hadronic
``band'' of width $\triangle \eta =\pm 0.9$ in the 
$(\eta_{\rm lab},\phi_{\rm lab})$-plane~\cite{rs}. 
The band reflects the isotropy
in the $I$-rest system. The total $E_T={\mathcal O}(20)$ GeV is large as
well as the multiplicity, $n_{\rm band}={\mathcal O}(25)$. Finally, 
there is a characteristic flavour flow: All (light) flavours are
democratically represented~\cite{th} in the final state. Therefore,
strongly enhanced rates of $K^{0}$'s and $\mu $'s 
(from strange and charm$^{\ref{charm}}$ decays) 
in the hadronic band represent crucial signatures 
for $I$-induced events.

\begin{figure}
\begin{center}
\epsfig{file=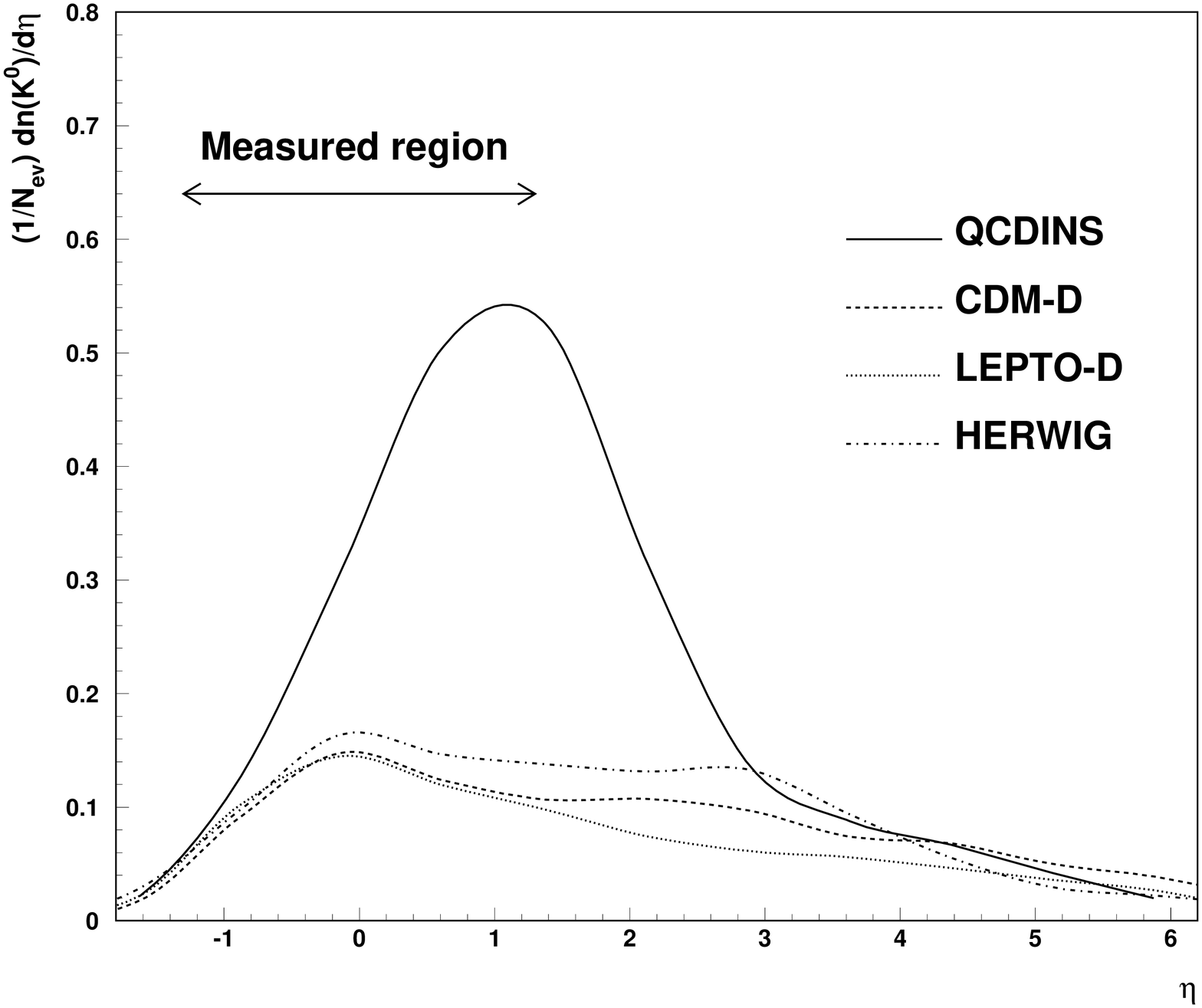,width=7.4cm}\hfill
\epsfig{file=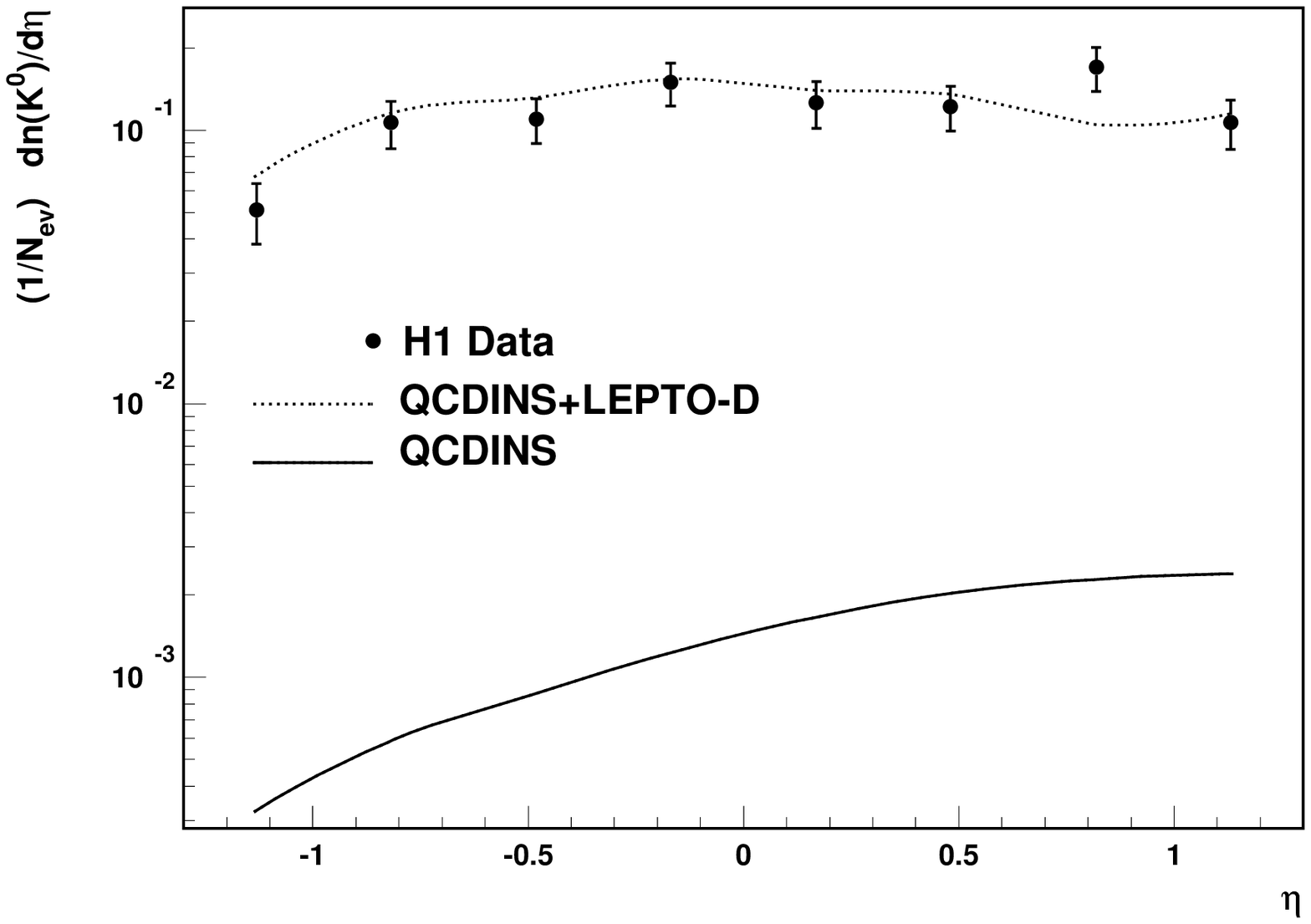,width=7.6cm}
\caption[dum]{\label{fkflows}
Numbers of $K^{0}$ mesons per event with $0.25<p_{T}^{2}<4.5\ {\rm GeV}^{2}$
as a function of the HERA-lab pseudorapidity $\eta$ for the kinematic
regime $10<Q^{2}<70\ {\rm GeV}^{2}$, $10^{-3}<x_{\rm Bj}<10^{-2}$, 
$0.1<y_{\rm Bj}<0.6$.
Left: Predictions from various models of (non-diffractive) DIS (broken lines) 
and from QCDINS (continuous line) for $I$-induced events only.
Right: H1 data (points) with the fit result (dotted line) and the fraction 
of instanton induced events (solid line)), $f=0.006$, obtained as described 
in the text. Both figures taken from Ref.~\cite{H1}.
}
\end{center}
\end{figure}

\subsection{A First Search for $I$-Induced Events at HERA}

If a significant proportion of DIS events at HERA were induced by
QCD-instantons, large changes in the strangeness composition of
the hadronic final state would be expected in the ``band'' region, as
illustrated in Fig.~\ref{fkflows} (left). 
This feature of $I$-induced events has been exploited by the
H1 collaboration~\cite{H1} who searched for an excess in the 
$K^{0}$ rate in the barrel of H1.

They fitted their measured rate of $K^{0}$ production, shown as a function
of the HERA-lab pseudorapidity $\eta$ in Fig.~\ref{fkflows} (right), 
with various DIS models, allowing 
for a fraction $f$ of $I$-induced events 
(see Fig.~\ref{fkflows} (left)). The largest value of $f$ was obtained 
using QCDINS with LEPTO~\cite{lepto} and was $f=0.006\pm 0.030$. 
The values $f$ found with different
DIS models were all found to be consistend with being zero.    
Hence, a 95\% confidence level upper limit of 0.9 nb was placed~\cite{H1} 
on the $I$-induced cross-section at HERA in the kinematical region considered
(c.\,f. Fig.~\ref{fkflows}). Note that this limit is 
less than an order of magnitude above our (preliminary) theoretical 
estimate presented in Fig.~\ref{fsigmacut}, which even contains stronger
additional cuts (in particular $\xpr \ge \xpr_{\rm min}= 0.2$). 

\subsection{Improving the Search Strategies}

\begin{figure}
\begin{center}
\epsfig{file=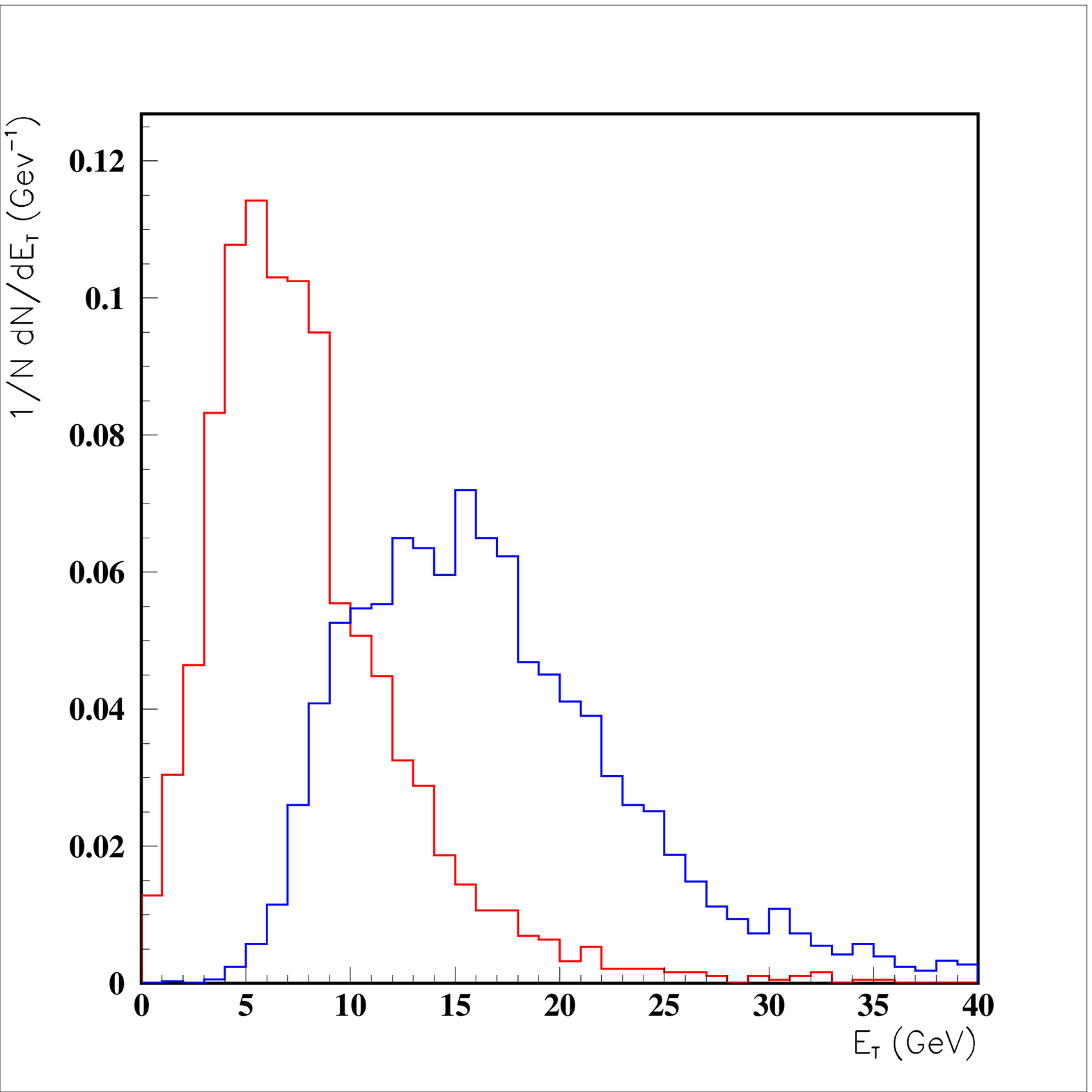,width=7.5cm,height=7cm}\hfill
\epsfig{file=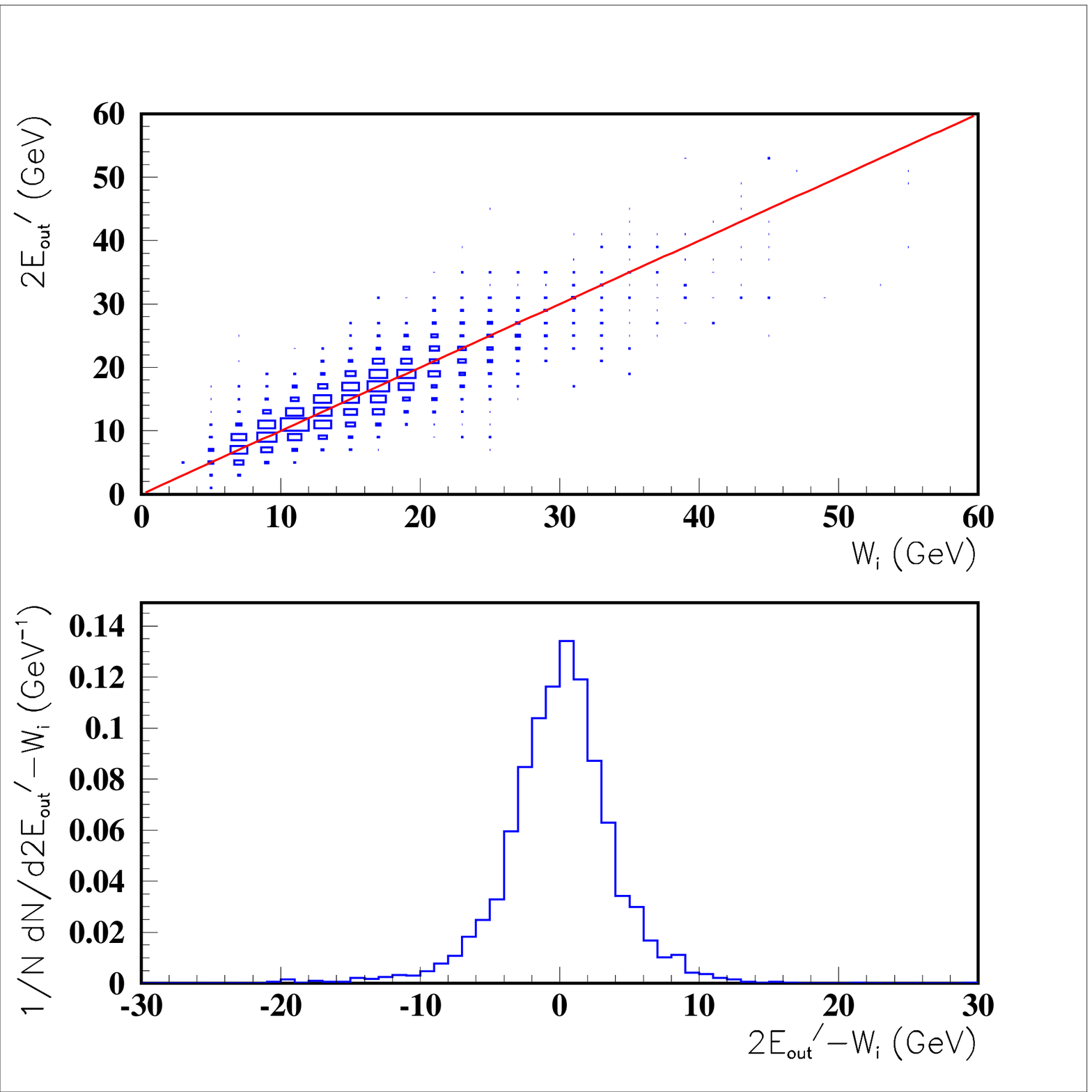,width=7.5cm,height=7cm}
\caption[dum]{\label{fetout}
Left: $E_{T}$ distributions in the $\gamma^{\ast} p$ c.m. system for 
normal DIS (left) and $I$-induced events (right). 
Right: Correlation between $2\,E_{\rm out}$ and the $I$-subprocess c.\,m. 
energy $W_I\equiv \sqrt{s^\prime}$. The primes indicate
additional cuts in $\eta$ to minimize next-to-leading-order perturbative
QCD effects. Both plots refer to the kinematic region 
$0.001<x_{\rm Bj}<0.01$, $0.1<y_{\rm Bj}<0.6$ and $20 < Q^2 < 70$ GeV$^2$.  
 }
\end{center}
\end{figure}

\begin{figure} 
\begin{center}
\epsfig{file=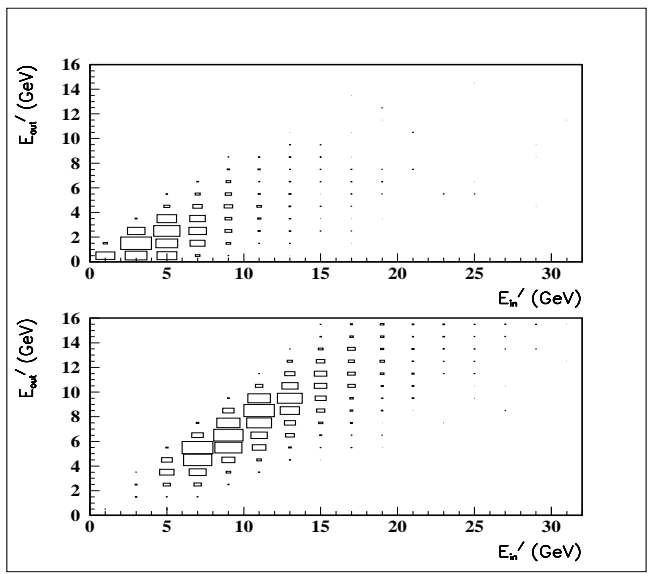,width=8.5cm,height=7.5cm}
\caption[dum]{\label{feineout} $E_{\rm out}$ vs. $E_{\rm in}$ 
distributions in the $\gamma^{\ast}$-proton c.m. system for normal DIS
events (top) and $I$-induced events (bottom) in the kinematic region 
$0.001<x<0.01$, $0.1<y<0.6$ and $20 < Q^2 < 70$ GeV$^2$. The primes indicate
additional cuts in $\eta$ to minimize next-to-leading-order perturbative
QCD effects. 
 }
\end{center}
\end{figure}

Let us finally mention some recent efforts~\cite{ggmrs} 
to improve the sensitivity to $I$-induced events by adding-in 
characteristic information on the {\it event shapes}.
The first step consists in boosting to the $\gamma^{\ast} p$ c.m. 
system and looking for events with high $E_{T}$ 
(c.f. Fig.~\ref{fetout} (left)).
We note that in this system (1+1) and (2+1) jet\footnote{As usual in DIS, 
the $+1$ refers to the proton remnant.} perturbative
QCD processes deposit their energy predominantly in a {\it plane} passing
through the $\gamma^{\ast} p$ direction. 
In contrast, the energies from $I$-induced events are always distributed 
much more {\it spherically} (isotropy in the $I$-rest system!).
Therefore, one may substantially reduce the normal DIS background 
by minimizing (on an event-by-event basis) the quantity
\begin{equation}
E_{\rm out}=\min_{\hat i}\,\sum_{k}^{n} \mid \vec{p}_{k}\cdot {\hat i} \mid\, ,
\end{equation}
by choice of the unit vector ${\hat i}$, normal to the 
$\gamma^{\ast} p$ direction. For standard 
(2+1) jet events from boson gluon fusion, $E_{\rm out}$ is then only of 
order of the jet widths. In contrast, for
$I$-induced events, $E_{\rm out}\simeq\sqrt{s^{\prime}}/2$ is large 
(see Fig.~\ref{fetout} (right)). Note, that the peaking of the $E_{T}$ and
$\sqrt{s^{\prime}}\equiv W_{I}$ distributions in Fig.~\ref{fetout} around
$15$ GeV directly reflects the shape of $\sigma_{q^{\ast}g}$ as function
of $Q^{\prime 2}\propto s^{\prime}$ (c.\,f. Fig.~\ref{falpha} (right)).
The quantitative results from the Monte-Carlo simulation, subject to additional
cuts in $\eta$ which are to minimize higher-order perturbative QCD effects,
are displayed in Fig.~\ref{feineout}. They fully confirm the 
qualitative expectations. The power of an  $I$-induced event selection 
based upon event shape is certainly apparent.

The production of large numbers of partons
in the $I$-subprocess also leads to large charged particle
multiplicities in the hadronic final state (c.f. Fig.~\ref{flego}). This 
is of particular interest for an instanton search, if the HERA
detectors can extend their multiplicity measurements to include
the region covered by the forward trackers. 

Apart from the enhancement of the $K^{0}$ production rate, 
a second feature associated with the flavour democracy of 
$I$-induced events is the large number of muons they contain. 
These result largely
from the decay of charmed particles$^{\ref{charm}}$. Unfortunately, their 
energies are rather low, with the transverse momenta of the muons in the 
laboratory frame being typically less than $1.5$ GeV. Hence, 
their detection is a challenging task for the experimenters.

Another challenging task is to study, how well the $I$-subprocess variables 
$s^\prime , x^\prime$ may be reconstructed/restricted from the hadron momenta 
in the ``band'' and the current jet. This would enable us to compare with
the theoretical estimates of the production rates which, as explained in
Sect.~\ref{subsec:sigqg}, are only available in the range $\xpr\gwig 0.2$,
$Q^{\prime}\gwig 5$ GeV.  

\section{Conclusions}

The experimental discovery of QCD-instanton induced events would 
clearly be of basic significance, since they correspond to a novel, 
non-perturbative manifestation of QCD. 
In addition, they would also provide valuable indirect information about  
$(B+L)$-violation in the multi-TeV region, induced by electro-weak
instantons.   

In this review, we have presented a status report of our 
broad and systematic study of QCD-instanton induced processes
in deep-inelastic scattering~\cite{rs,grs,rs1,ggmrs,mrs1,mrs2,grs2}.

We have emphasized that deep-inelastic scattering may be viewed as 
a distinguished process for studying manifestations of QCD-instantons,
since the available hard scale provides a dynamical infrared cutoff for the
instanton size~\cite{mrs1}. At high $Q^2$, instanton-induced 
{\it fixed-angle} processes may be reliably calculated, like in 
perturbative QCD. Thus, deep-inelastic scattering at HERA offers a unique 
window to explore footprints of QCD-instantons.  

Of great interest is, of course, a first 
estimate of the total instanton-induced cross-section at HERA.     
We pursued a complementary approach to Ref.~\cite{bb}, by 
working  out~\cite{mrs2} a momentum-space picture of the 
instanton-contribution to the parton structure functions and further 
inclusive observables.
In leading-order of the semi-classical approximation about the 
in\-stan\-ton-anti\-in\-stan\-ton configuration and at large $Q^2$, the 
instanton-contribution to the gluon structure function 
${\mathcal F}_{2\, g}$ has the form of a 
convolution of a ``splitting function'' in the instanton-background
with a total cross-section for the instanton-induced $q^\ast g$ subprocess.  
The latter contains the essential instanton dynamics.
We presented a preliminary theoretical estimate of the total 
instanton-induced cross-section for deep-inelastic scattering at HERA, 
subject to appropriate kinematical cuts. Notably, the Bjorken scaling 
variable of the  $q^\ast g$ subprocess has to be restricted from below, in 
order to retain theoretical control.
Being in the  ${\mathcal O}(1-100)$ pb range, the resulting cross-section is 
surprisingly large.
Due to inherent uncertainties associated with the renormalization- and 
factorization-scale dependencies, which are presently being 
investigated~\cite{mrs2}, this cross-section is, however,  still to be 
considered preliminary.

On the phenomenological side, we systematically explored the discovery 
potential for instanton-induced events at HERA,
by studying the characteristics of the final state: A current-quark
jet along with a densely populated hadronic ``band'' of width 
$\triangle\eta =\pm 0.9$ in the $(\eta_{\rm lab},\phi_{\rm lab})$-plane,
reflecting the {\it isotropic} decay of an instanton-induced fireball.  
Characteristic features include the large total transverse energy,
$E_{T}={\mathcal O}(20)$ GeV, the large multiplicity, 
$n_{\rm band}={\mathcal O}(25)$, and the flavour-democratic production of
hadrons, leading in particular to abundant production of $K^{0}$ mesons.
At present, we consider the characterization of instanton-induced events by 
these final-state features to be more robust than predictions based on
cross-section estimates.

From a recent measurement of $K^{0}$ production in deep-inelastic scattering
at HERA, the H1 collaboration placed a first upper limit of $0.9$ nb 
(95\% confidence level) on the cross-section for instanton-induced events
in the kinematical region considered. This limit is less than an order of 
magnitude above our (preliminary) theoretical estimate, which even contains
stronger additional cuts. We have studied in detail, how 
a combination of event shape information with searches of
$K^{0}$ mesons, muons, and multiplicity cuts may 
further help to discriminate the QCD-instanton induced processes from the
standard perturbative QCD background.

\vspace{0.2cm}
\section*{Acknowledgements}
We would like to acknowledge helpful discussions with 
V. Braun, V. Rubakov and C. Wetterich.

\end{document}